\documentclass{emulateapj}

\newcommand\gsim{\,\lower3pt\hbox{$\sim$}\llap{\raise2pt\hbox{$>$}}\,}
\newcommand\lsim{\,\lower3pt\hbox{$\sim$}\llap{\raise2pt\hbox{$<$}}\,}

\usepackage{graphicx}
\usepackage{epsfig}
\usepackage{subfigure}
\usepackage{natbib}

\shortauthors{LUGAZ ET AL.}

\shorttitle{SIMULATION OF A CME FROM ANEMONE ACTIVE REGION 10798}

\begin{document}

%
%

\title{Numerical Investigation of a Coronal Mass Ejection from an Anemone Active Region: Reconnection and Deflection of the 2005 August 22 Eruption}

\author{N.\ Lugaz \altaffilmark{1, 2}, C.\ Downs \altaffilmark{2}, K.\ Shibata \altaffilmark{1}, I.~I.\ Roussev \altaffilmark{2}, A.\ Asai \altaffilmark{3}, T.~I.\ Gombosi \altaffilmark{4}}

\altaffiltext{1}{Kwasan Observatory, Kyoto University, Kyoto, Japan}
\altaffiltext{2}{Institute for Astronomy, University of Hawaii, Honolulu, HI, USA}
\altaffiltext{3}{Unit of Synergetic Studies for Space, Kyoto University, Kyoto, Japan}
\altaffiltext{4}{Center for Space Environment Modeling, University of Michigan, MI, USA}

%
%

\begin{abstract}
We present a numerical investigation of the coronal evolution of a coronal mass ejection (CME) on 2005 August 22 using a 3-D thermodynamics magnetohydrodynamic model, the SWMF. The source region of the eruption was anemone active region (AR) 10798, which emerged inside a coronal hole. We validate our modeled corona by producing synthetic extreme ultraviolet (EUV) images, which we compare to EIT images. We initiate the CME with an out-of-equilibrium flux rope with an orientation and chirality chosen in agreement with observations of a H$\alpha$ filament. During the eruption, one footpoint of the flux rope reconnects with streamer magnetic field lines and with open field lines from the adjacent coronal hole. It yields an eruption which has a mix of closed and open twisted field lines  due to interchange reconnection and only one footpoint line-tied to the source region. Even with the large-scale reconnection, we find no evidence of strong rotation of the CME as it propagates. We study the CME deflection and find that the effect of the Lorentz force is a deflection of the CME by about 3$^\circ$~/~R$_\odot$ towards the East during the first 30 minutes of the propagation. We also produce coronagraphic and EUV images of the CME, which we compare with real images, identifying a dimming region associated with the reconnection process.  We discuss the implication of our results for the arrival at Earth of CMEs originating from the limb and for models to explain the presence of open field lines in magnetic clouds.

\end{abstract}

\section{INTRODUCTION} \label{intro}
Coronal mass ejections (CMEs) are one of the leading causes of space weather, especially when they have organized southward directed magnetic field. Therefore, it is of great importance for space weather forecasting to understand how their direction of propagation and their orientation at 1~AU relate to properties on the solar disk (location of source region, orientation of the polarity inversion line, etc...). The deflection of CMEs in the latitudinal direction has been observed and reported since the launch of the first space coronagraphs in the 1970s and 1980s. For example, \citet{MacQueen:1986} reported an average deflection of 2.2$^\circ$ towards the equator for 29 CMEs during solar minimum (1973--1974), while they found no systematic deflection for 19 CMEs during solar maximum (1980). With the launch of the {\it Solar-Terrestrial Relations Observatory} (STEREO) in 2006, there has been a renewed focus on the deflection of CMEs in the corona and the heliosphere, thanks to stereoscopic measurements from the two spacecraft \citep[e.g., see:][]{Kilpua:2009a,Liu:2010b, Byrne:2010}.

The presence of coronal holes is known to affect the direction of propagation of CMEs \citep[]{Plunkett:2001, Gopalswamy:2009}. However, the exact cause of this influence is uncertain. \citet{Cremades:2006} and \citet{Gopalswamy:2009} have used an {\it ad hoc}, fictitious force exercised by coronal holes on CMEs, which is directly dependent on the coronal hole area and the distance of the CME source region from the coronal hole. This force might be due to a ``pushing'' of the CME by the fast wind from the coronal hole \citep[]{Wang:2004}. \citet{Plunkett:2001} proposed that the effect of coronal holes is due to strong magnetic fields, \citet{Filippov:2001} proposed that the non-radial motion is due to the guiding action of the coronal magnetic field, while \citet{Aulanier:2010} and \citet{Shen:2011} recently proposed that magnetic pressure gradient acting on a CME results in a net force directed along the gradient of the magnetic pressure. Obviously, magnetic forces, such as the Lorentz force could also cause the coronal hole to exercise a force on the CME. Magnetic forces are expected to affect strongly the evolution and propagation of a CME in the corona while the effect of the fast solar wind from the coronal hole would continue into the heliosphere and be more gradual. 

Similar effects may result in the rotation of an eruption. One of the leading causes of CME rotation is the kink instability \citep[]{Torok:2003} but recent numerical efforts have focused on other sources of rotation, such as reconnection with the background magnetic field \citep[]{Cohen:2010}, the effect of the Lorentz force \citep[]{Isenberg:2007, Shiota:2010} and its association with shearing motions \citep[]{Lynch:2009}. These numerical works point towards a near universal rotation for CMEs not initially aligned with the heliospheric current sheet (HCS). While statistically speaking, CMEs rotate towards the HCS \citep[]{Yurchyshyn:2008}, it appears from observations that some CMEs do not rotate, even though they are originally not aligned with the HCS \citep[see some cases in][]{Yurchyshyn:2008, Yurchyshyn:2009}. 

Eruptions from active regions which are inside a coronal hole present a perfect configuration to study the effect of coronal holes on CMEs. For example, the fictitious force of \citet{Cremades:2006} is inversely proportional to the square of the distance between the source region and the coronal hole, and it results in a infinite force for a CME originating from such an active region!
Active regions can form inside coronal holes and, then, they often develop into anemone active regions \citep[]{Shibata:1994, Asai:2008}. Anemone active regions were originally observed in soft X-ray and named X-ray fountains \citep[]{Sheeley:1975} due to the structure looking like a fountain with loops emerging in all directions from the center. Almost identical structures are observed in extreme ultraviolet (EUV) and in chromospheric lines \citep[]{Shibata:2007}. In a survey of active regions observed by the soft X-ray telescope \citep[SXT, see:][]{Tsuneta:1991} onboard {\it Yohkoh} \citep[]{Ogawara:1991}, \citet{Asai:2008} found that as many as 25$\%$ of active regions observed in 1991--1992 appeared as anemone active regions at one time during their evolution, and that there is a near equivalence between an active region having an anemone structure and being inside a coronal hole.

Anemone active regions have been primarily studied as the source of X-ray jets since the emergence of a bipolar active region inside unipolar open magnetic fields naturally yield such phenomena \citep[]{Shibata:1994, Vourlidas:1996}. Of greater importance for space weather are the instances of CMEs and filament eruptions from anemone active regions \citep[]{Chertok:2002,Asai:2009, Baker:2009}. \citet{Liu:2007} found that eruptions from active regions inside open magnetic field (which contain anemone active regions) are statistically faster on average than other eruptions (even originating from under the heliospheric current sheet). Most authors have explained this result by invoking  two facts. Since the eruption originates from a region of low-density, fast solar wind, the fast wind can ``push'' the CMEs to a faster speed. Closed field in the low corona can hinder the eruption of the CME because of the Lorentz force \citep[see for example the model of][]{Chen:1996} and in the absence of closed field lines, the CME can erupt without being strongly decelerated. 

In the present work, we investigate the effect of the particular magnetic structure of anemone active regions on the evolution and coronal propagation of a CME. We focus on the first of two eruptions in 2005 August 22 from active region NOAA 10798. This CME was associated with the eruption of a southward-directed H$\alpha$ filament and reached a speed of about 1200~km~s$^{-1}$ \citep[for a complete overview of the observations, see][]{Asai:2009}. The second eruption was faster ($\sim$ 2400~km~s$^{-1}$) and it occurred about 16 hours after the first one from the same active region. The eruption of a preceding CME from the same active region is expected to significantly change the magnetic topology of the source region \citep[]{Lugaz:2010a} and also modify the background solar wind into which the second CME propagates \citep[]{Lugaz:2005b, Lugaz:2007}, which would complicate a numerical investigation of the second CME. For this reason, we focus on the first CME. 
Our investigation is based on a numerical magnetohydrodynamic (MHD) model part of the space weather modeling framework \citep[SWMF, see:][]{Toth:2005}. In order to validate our simulation with EUV observations and to reproduce more accurately the lower part of the corona, we use the low corona (LC) model recently developed by \citet{Downs:2010}, which captures the energy balance of the coronal plasma.

The organization of the paper is as follows: in Section~\ref{sec:models}, we briefly summarize the main features of the LC and flux rope models important for this study.  Then, we discuss the magnetic topology of the anemone active region before presenting a comparison of the pre-event corona as observed in EUV with the modeled corona in synthetic EUV. In Section~\ref{sec:TD}, we follow the initial phases of the eruption and explain its interaction with the adjacent magnetic flux systems. In this section, we also discuss the CME aspect in real and synthetic coronagraphic images. We study the CME rotation, deformation and deflection in Section~\ref{sec:rotation}. We discuss our results and conclude in Section~\ref{sec:conclusion}.


\section{NUMERICAL MODELS AND PRE-EVENT MODELING} \label{sec:models}
\subsection{Low Corona Model}
The simulation is done using the Low Corona (LC) component of the
SWMF. The LC model includes radiative losses, field-aligned electron heat condition and empirical heating terms in the energy equation as detailed in \citet{Downs:2010}. It has been recently used to study the nature of the extreme ultraviolet (EUV) wave observed by the two {\it STEREO} spacecraft in 2008 March 25 \citep[]{Downs:2011}. The initial magnetic field and boundary conditions are provided by a finite difference solution of the potential field source surface method \citep[PFSSM:][]{Altschuler:1977} for the synoptic magnetogram of Carrington rotation 2033 as observed by the {\it SOHO}/MDI instrument \citep[]{Toth:2011}. As noted in \citet{Asai:2009}, AR 10798 was rapidly evolving as it crossed the Sun central meridian and therefore, the synoptic map does not fully capture the complexity of the active region as revealed by observations on August 21 and 22. However, on these two days, the AR was already too close to the eastern limb of the Sun, so that the daily line-of-sight magnetogram (shown in the left panel of Figure~1) cannot be adequately used to constrain our model. As discussed below, even using the synoptic map, the MHD model successfully reproduces the important features of the anemone active region. 

As described in \citet{Downs:2010}, we followed the previous works of \citet{Lionello:2001} and \citet{Lionello:2009} to widen the transition region by modifying the ratio of heat conduction to radiative cooling at temperatures below 300,000~K. The heating model and boundary conditions are the same as the ones described in \citet{Downs:2011}, i.e. we use the chromospheric boundary conditions. 
The grid is spherical with non-uniform radial scales near the transition region (minimum uniform $dr \sim$ 230~km) with a resolution on the solar surface of $1.4^\circ \times 1.4^\circ$. In addition, we further refine around AR 10798 by 2 levels of refinement as follows. The solar surface is refined twice in a box of about 20$^\circ$ latitude and $40^\circ$ longitude around the active region over a height of 0.4~R$_\odot$ and the region around the top of the anemone structure (null point) is refined by an additional factor of 2 in a twice smaller box centered at 1.15~R$_\odot$, resulting in a smallest angular resolution of $0.35^\circ \times 0.35^\circ$. The grid around AR~10798 is shown in the top left panel of Figure~2.

 \begin{figure*}[t]
\begin{minipage}[]{1.0\linewidth}
\begin{center}
{\includegraphics*[width=10.5cm]{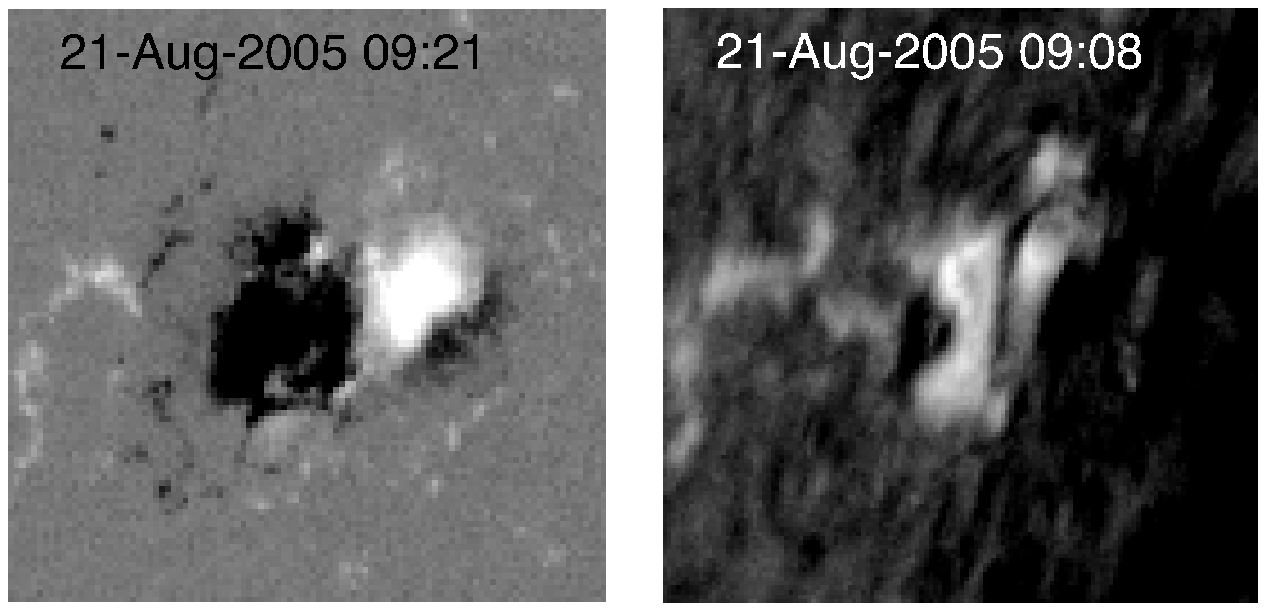}}
{\includegraphics*[width=5.5cm]{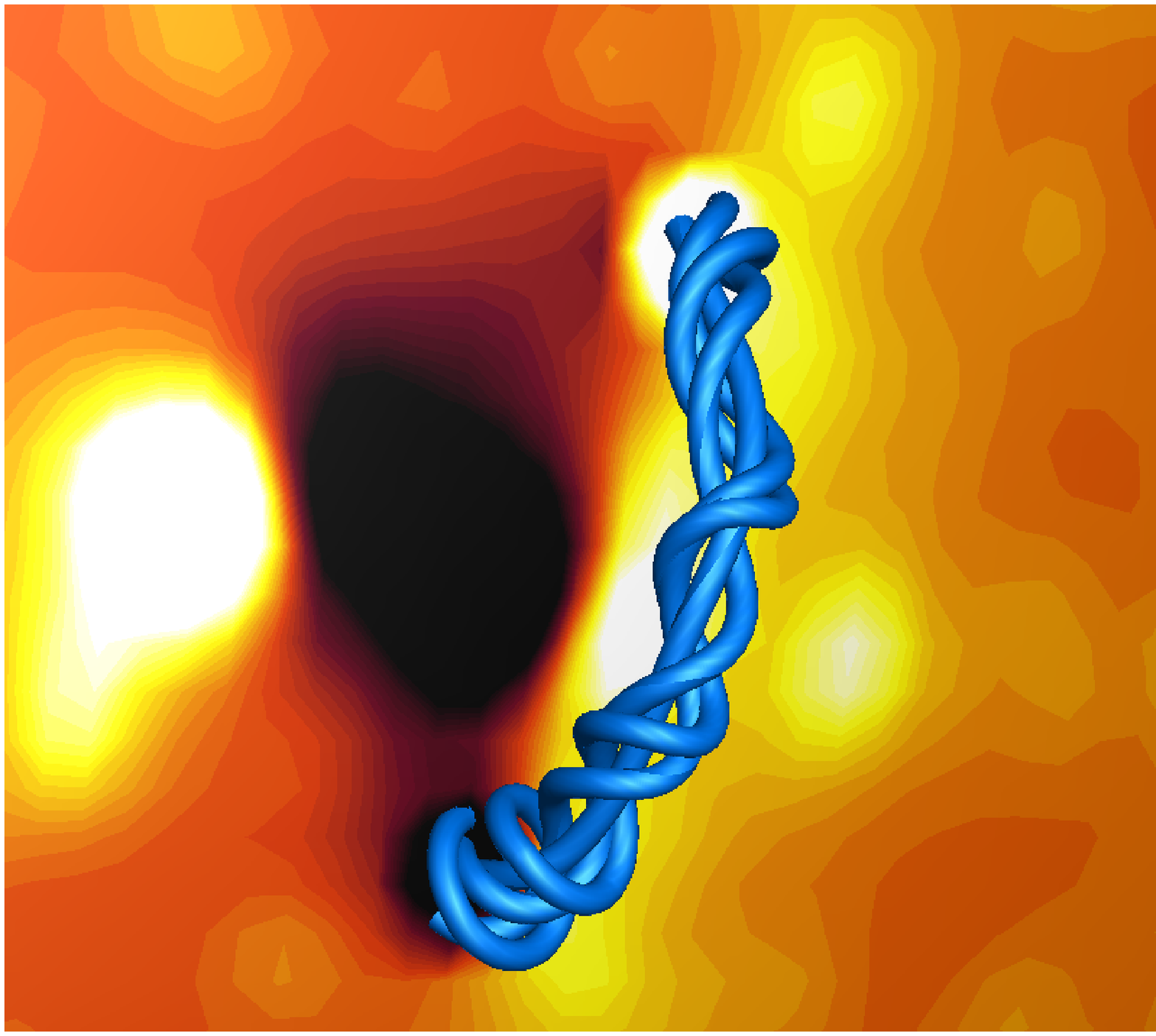}} 
\end{center}
\end{minipage}\hfill
\caption{Initial configuration of the filament. Left: SoHO/MDI magnetogram. Middle: H$\alpha$ image of the filament taken at the Observatoire de Paris--Meudon. Right: MHD simulation at time $t=0$ with the background magnetic field on the solar surface and the flux rope shown with blue field lines. In all three views, the images show approximatively 200 arcsec $\times$ 200 arcsec.}
\end{figure*}

This version of the LC model includes a uniform Parker-like solar wind model but not a bimodal solar wind model as in the models of \citet{Cohen:2007} and \citet{Holst:2010}. Here, the solar wind accelerates and becomes faster than the fast magnetosonic speed between 6 and 14~R$_\odot$ and reaches a value of 300--350~km~s$^{-1}$ at the outer boundary of the domain (24~R$_\odot$). While it is not realistic for the fast wind originating from coronal holes, it is the approximate speed and acceleration profile expected for the slow solar wind. We only focus on the coronal evolution of the CME, therefore resolving a bimodal solar wind is not critical to our study. Since the solar wind is uniform, we can expect a negligible amount of deflection and rotation due to hydrodynamical effect (velocity shear, non-uniform drag). It allows us to focus solely on the effect of magnetic forces and reconnection on the CME evolution.
Compared to previous studies of the evolution of CMEs in the corona with the SWMF \citep[]{Lugaz:2007, Cohen:2010, Evans:2011}, using the LC model has a number of advantages. The inclusion of the thermodynamics makes the treatment of the lower corona more realistic. Using the chromospheric boundary conditions, the density and temperature of the plasma are based solely on the magnetic topology and the particular choice of heating function. Also, as discussed in \citet{Downs:2010}, it results in a less potential steady-state solution for the background coronal magnetic field. Last, it gives us the opportunity to validate our model with a direct comparison to extreme ultraviolet (EUV) observations.

\subsection{CME Model}
To model the CME, we employ a semi-circular flux rope prescribed by a given total toroidal current, as in the models by \citet{Titov:1999} and \citet{Roussev:2003a}.  An azimuthal current is also added at the surface of the flux rope in order to construct a force-free magnetic field inside the flux rope. The toroidal current is largely dominant, creating a very highly twisted flux rope. A more complete description of this implementation of the flux rope model can be found in \citet{Lugaz:2007} and \citet{Evans:2011}. Because in this version of the \citet{Titov:1999} flux rope, there is neither sub-surface line current nor sub-surface magnetic charges to generate strapping field for the flux rope in the corona, the flux rope is not expected to rotate as found for example in \citet{Isenberg:2007}. Additionally, the flux rope is constructed so that it is not kink unstable. The flux rope solution once superimposed onto the coronal magnetic field leads to an immediate eruption because of force imbalance with the ambient magnetic field.

 \begin{figure*}[t]
\begin{minipage}[]{1.0\linewidth}
\begin{center}
{\includegraphics*[width=6.8cm]{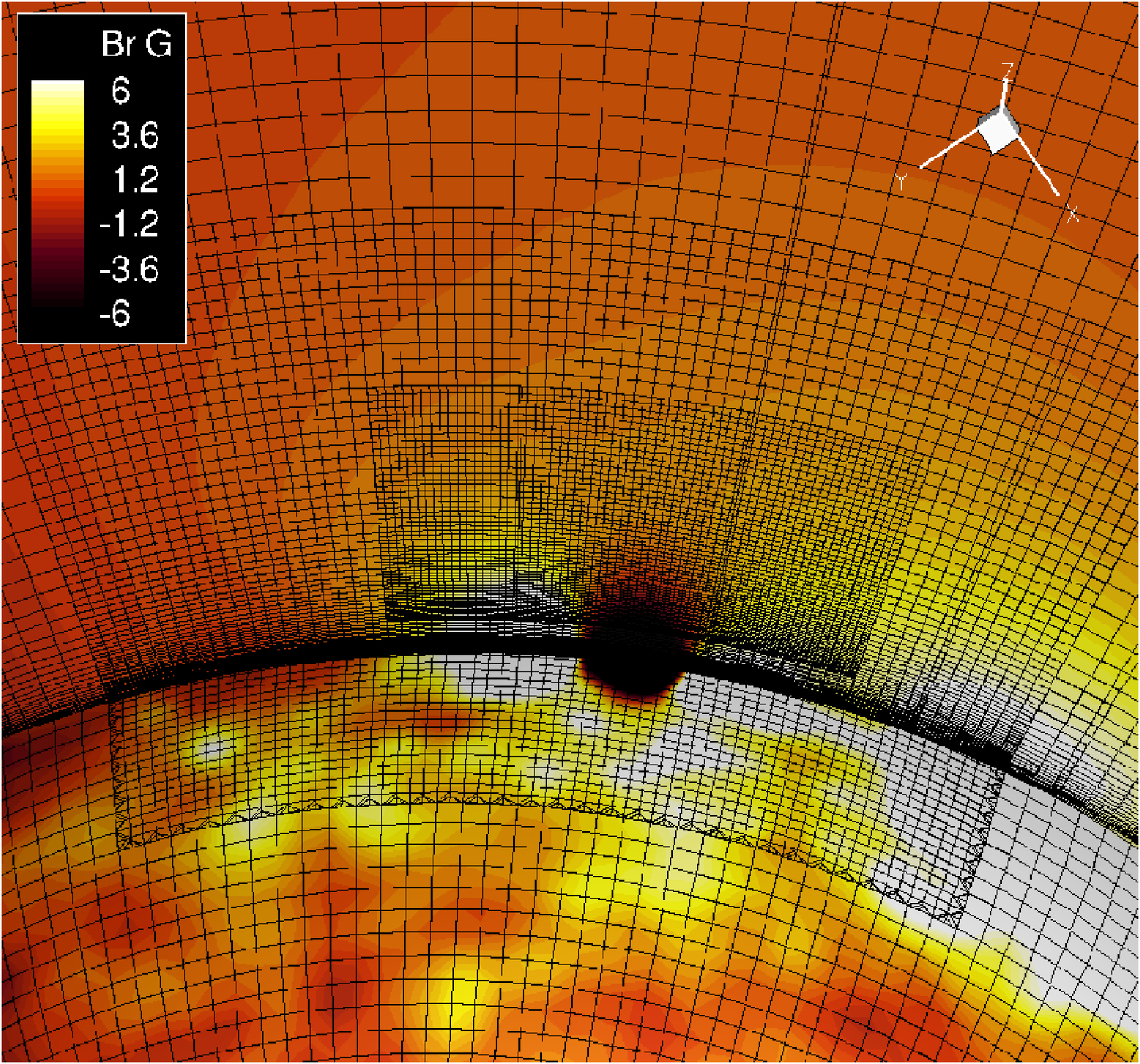}}
{\includegraphics*[width=7cm]{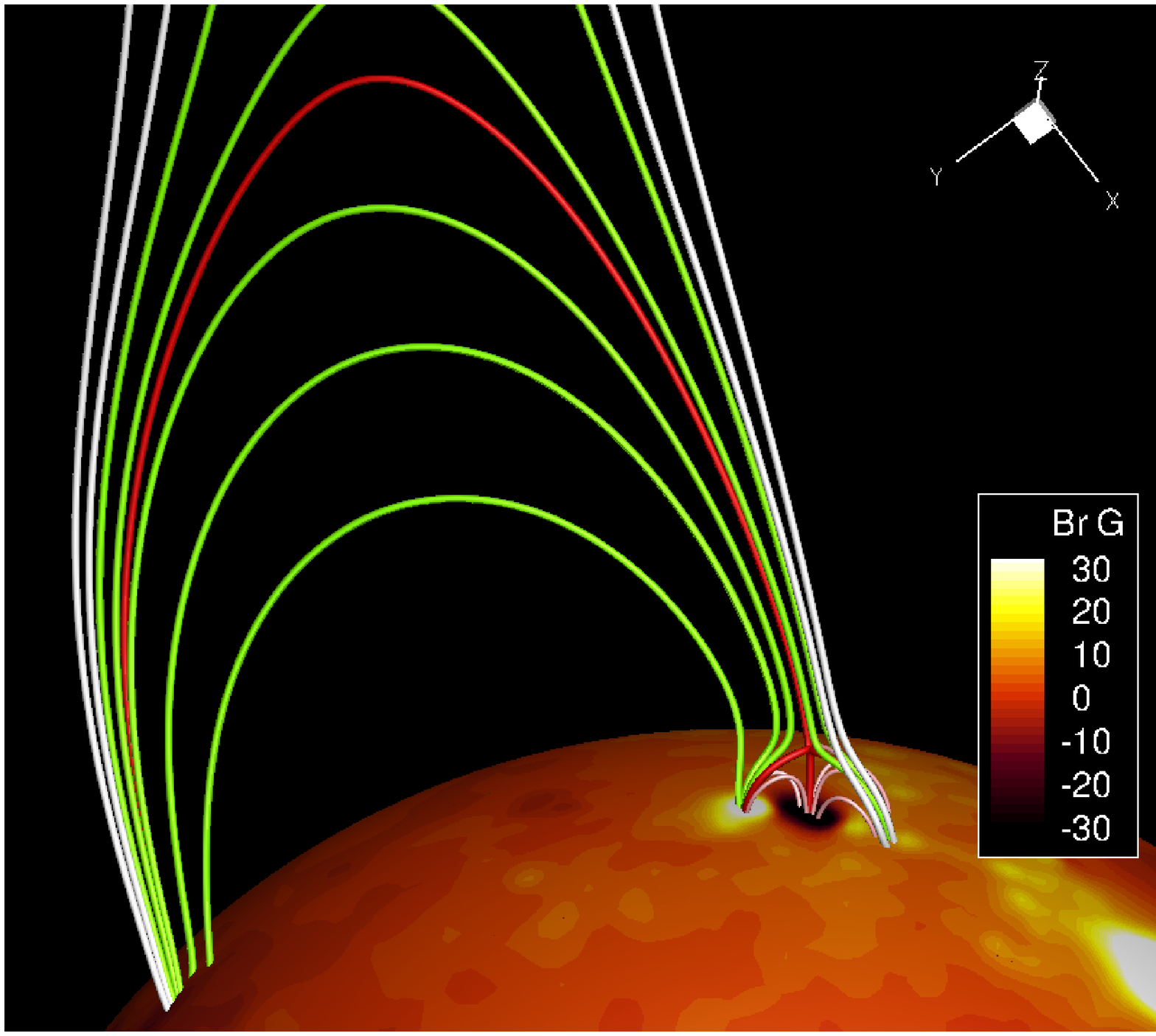}}\\
{\includegraphics*[width=7cm]{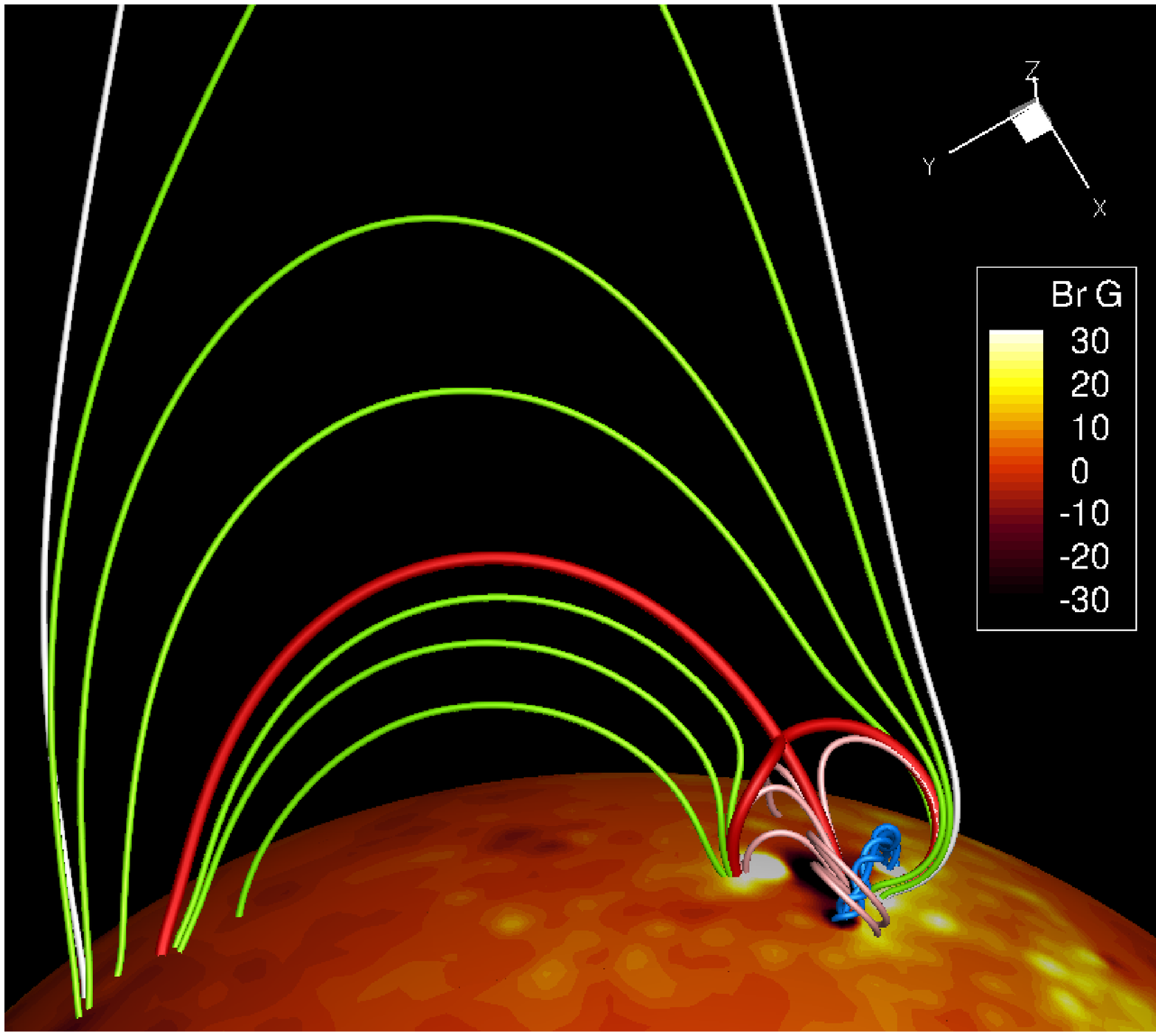}} 
{\includegraphics*[width=7cm]{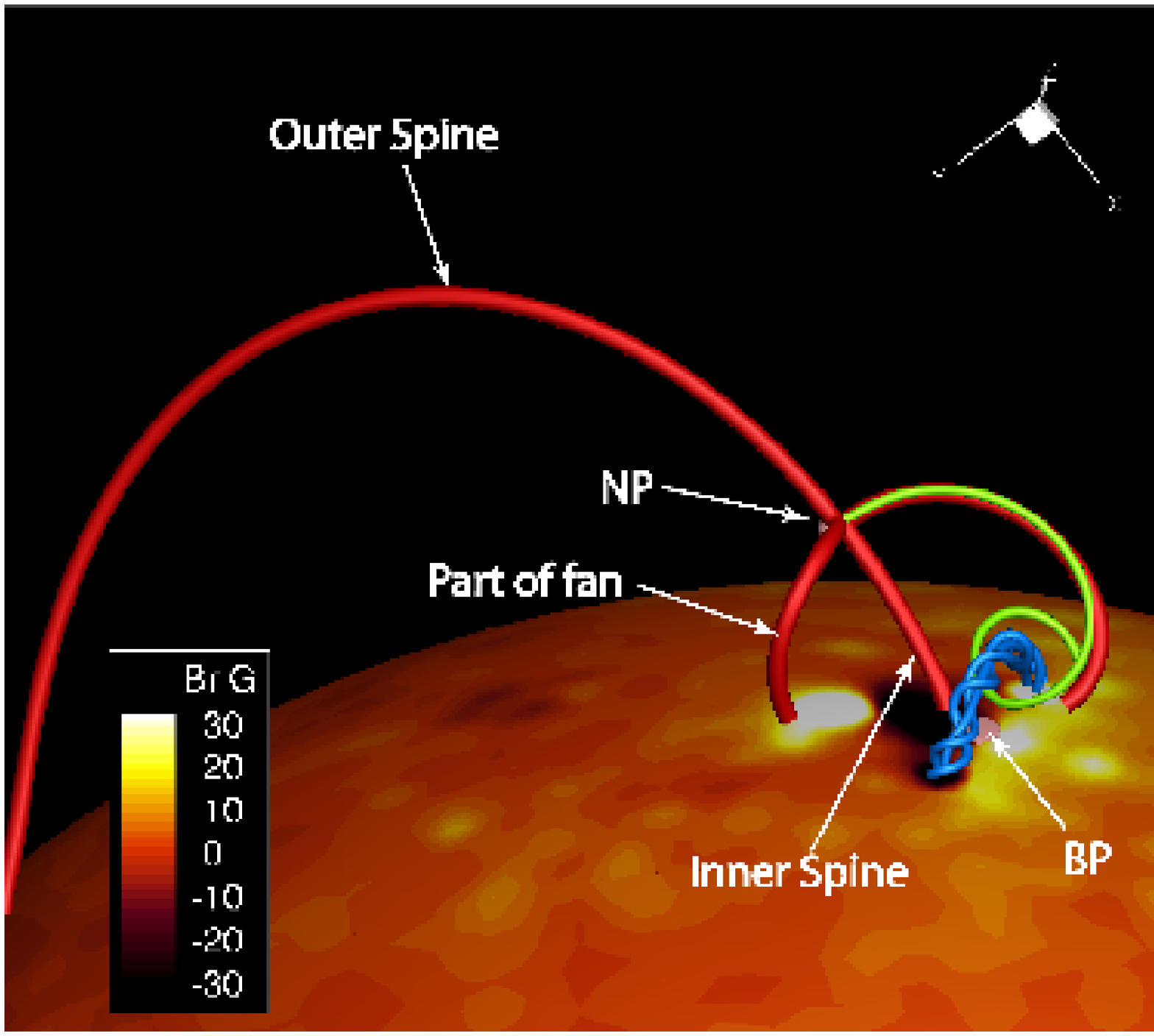}} 
\end{center}
\end{minipage}\hfill
\caption{Initial configuration of the corona around AR 10798 from the simulation with the grid structure (top left), the anemone structure of the active region before the inclusion of the flux rope (top right) and after (bottom panels).
White, green, pink and blue field lines show open, streamer, active region and flux rope magnetic field, respectively. The solar surface is color-coded with the radial magnetic field strength. The bottom right panel shows the zoomed version of the flux rope as it is added to the solar surface with a view of the bald patch (BP) and the null point (NP) as pink isosurfaces. In red, we show the inner and outer spines of the system as well as one of the fieldlines of the fan surface.}
\end{figure*}

A filament was observed on August 21 (see middle panel of Figure~1) and the eruption of the northern section of the filament was the cause of the studied eruption \citep[]{Asai:2009}. We place the flux rope at a position and with an orientation in agreement with the observation of the filament (see Figure~1). The flux rope (and the line current inside) makes an angle of $-15^\circ$ with the $-z$ axis (the Sun rotation axis). It results in a right-handed flux rope with an axial southward magnetic field, in agreement with the observations of the filament in H$\alpha$ and with the reconstruction of the associated ejecta at 1~AU \citep[]{Asai:2009}. The flux rope is making an angle of about 20$^\circ$ with the polarity inversion line of AR 10798 with its negative (resp. positive) footprint in the main part of the negative (resp. positive) polarity spot of the AR. The direction of the axial field of the flux rope is chosen in agreement with the observations of a southward (sinistral) filament \citep[]{Asai:2009} which is also consistent with the prefilament structure  \citep[see Figure~4 from][]{Asai:2009}. The orientation of the axis of the flux rope is consistent with the observations of a structure oriented from the northwest to the southeast. The exact orientation is chosen so that the positive (resp. negative) footprint of the flux rope are in a region of positive (resp. negative) polarity, while the axis of the flux rope remains close to the polarity inversion line. It should be noted that in our MHD model, because we use a synoptic magnetogram, the positive polarity spot of the active region (on the west side of the negative spot) is not yet fully developed while the positive polarity on the east side is stronger that it is on August 21--22. The flux rope is chosen as right-handed to agree with the overall magnetic field in the overlying arcades once the direction of the axial field (sinistral) of the flux rope is chosen. This is also consistent with what was reported in \citet{Asai:2009}. Finally, the amount of twist is determined by the prescribed toroidal current. The current is set by a trial-and-error procedure to match the coronal speed of the CME as observed by LASCO (1200 km~s$^{-1}$). Because of the way our flux rope is created (with a dominant toroidal current), the chosen value of the current results in a flux rope with a much larger amount of twist as compared to what is derived from observations \citep[see review by][]{MacKay:2010}. 

Previous time-dependent simulations with the LC model were performed with the CME initiation model of \citet{Roussev:2007}. Here, we use, for the first time, a flux rope model in the LC model because the presence of a filament strongly suggests that a flux rope was present prior to the eruption. This CME model has been used before to study the evolution of CME in the corona \citep[]{Lugaz:2007, Cohen:2010, Evans:2011}. A view of the flux rope at time $t =0$ as it is superposed onto the steady-state corona is shown with blue field lines in the bottom panels of Figure~2.

 \begin{figure*}[t]
\begin{minipage}[]{1.0\linewidth}
\begin{center}
{\includegraphics*[width=15cm]{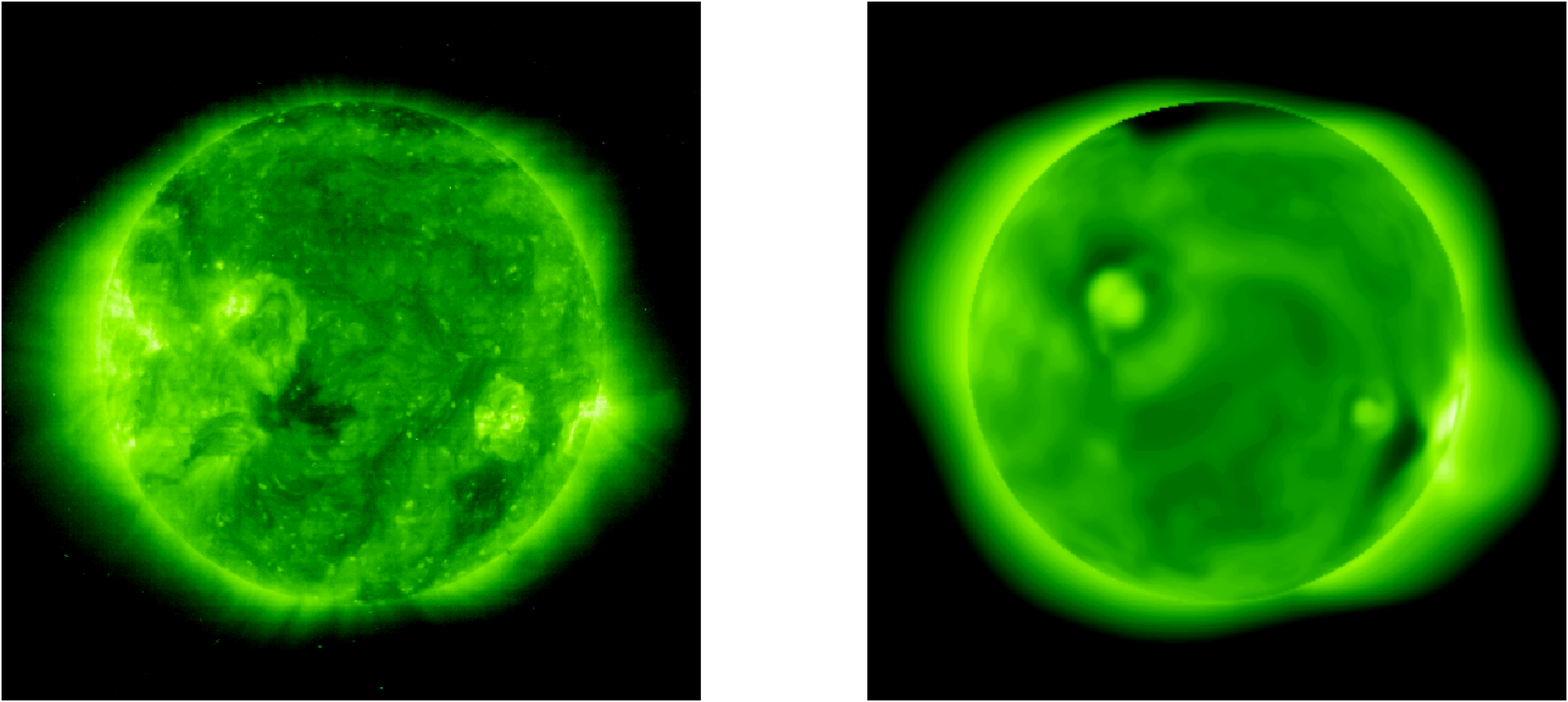}}
{\includegraphics*[width=15cm]{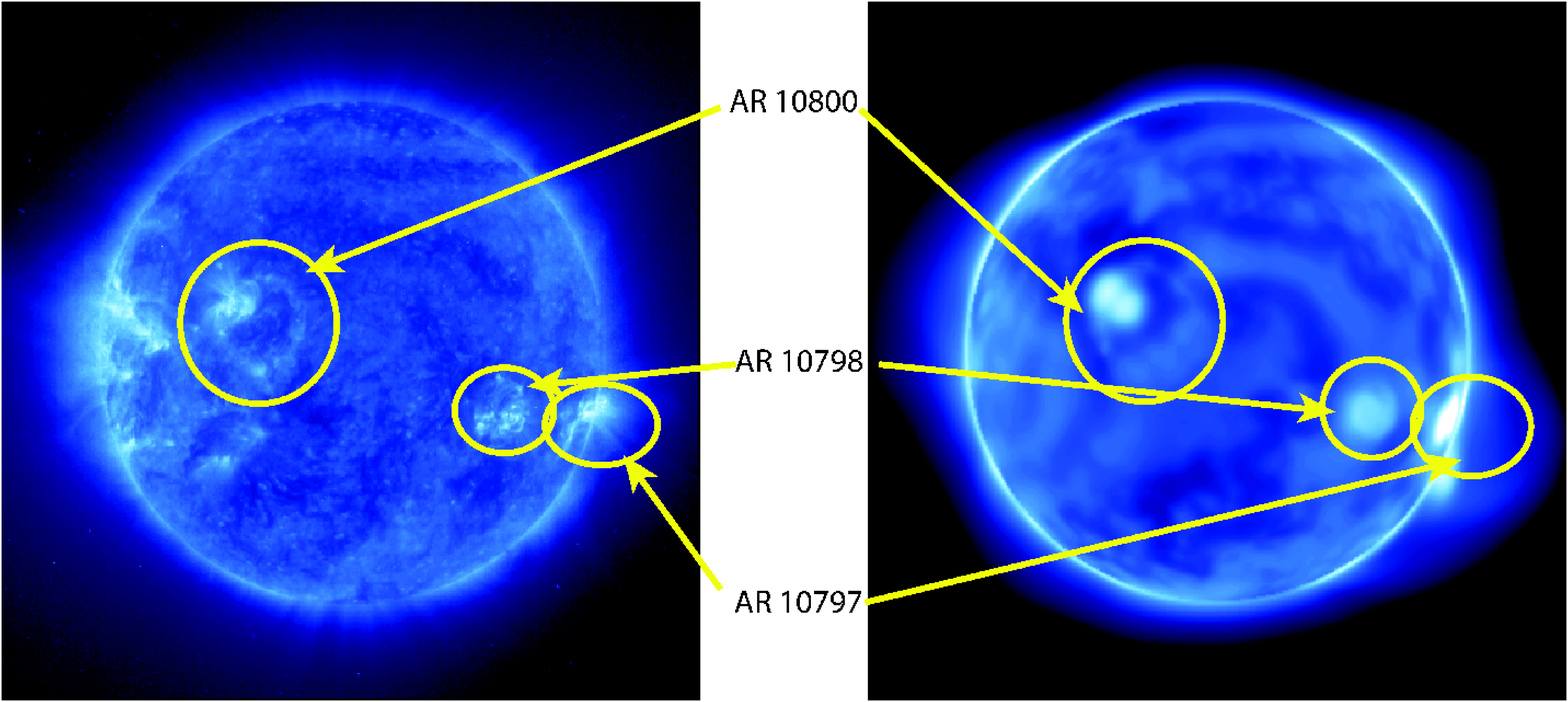}}

\end{center}
\end{minipage}\hfill
\caption{Real ({\it left}) and synthetic ({\it right}) EIT 195\AA~image (top) and 171\AA~image (bottom) of the corona 18 hours before the start of the first 2005 August 22 eruption and from our steady-state simulated corona. On the 171\AA~images, the main active regions (ARs) are pointed out in yellow ellipses.}
\end{figure*}

\subsection{Pre-event Topology}
The magnetic topology before the eruption is very important because it determines, to a great extent, the location of the reconnection of the flux rope with adjacent magnetic field structures during the eruption.
The steady-state corona around AR 10798 is relatively simple with: 1) a large-scale streamer connecting the active region with a region of the quiet Sun about 40$^\circ$ east of the active region, 2) open field lines with positive polarity, and, 3) active region closed field lines. We found no magnetic connectivity between AR 10798 and adjacent active regions, even though AR 10797 is only about 20--25$^\circ$ away. It is not surprising since AR 10798 is a reverse polarity active region (negative-positive) while AR 10797 follows the \citet{Hale:1919} polarity law (positive-negative). Therefore the positive parts of ARs 10798 and 10797 are next to each other and, moreover, they are embedded into positive polarity open magnetic fields from the coronal hole. This type of magnetic field configuration make it very unlikely that there was a direct magnetic connectivity between these two active regions. The topology is much simpler than other cases near solar maximum, where \citet{Roussev:2007} found connectivity between as many as three active regions via multiple null points and quasi-separatrix layers. 

There is one null point above the negative polarity part of AR 10798 separating the 3 flux systems as illustrated in the Figure~2: AR field (pink), streamer field (green) and open field (white). Because the negative polarity spot is surrounded by positive polarities, it develops into an anemone active region (see top right panel of Figure~2) as described before in \citet{Asai:2008} and \citet{Shibata:2007}. This null point is originally at a height of 0.09~R$_\odot$ above the solar surface. This value is probably lower than that on the Sun at the time of the eruption because we use a synoptic map of the Sun and AR 10798 was not fully developed. The existence of close field lines as part of a unipolar region has been reported before \citep[]{Chertok:2002}. 
The steady-state configuration is an embedded dipole, identical to that of the asymmetrical breakout model \citep[]{Lynch:2009} and similar to that of coronal jet studies \citep[e.g., see][]{Pariat:2009} but with a closed outer spine. Here, the inner and outer spines are initially closed and there is a simple fan surface around the negative polarity spot of the active region. The inner and outer spines as well as one of the lines of the fan surface are shown in red in Figure~2.
As we superpose the flux rope onto the active region, the null point is pushed by about 4$^\circ$ toward the north-east and by 0.04~R$_\odot$ upward, but the active region retains its anemone structure (see bottom panels of Figure~2). In addition, a bald patch forms below the flux rope (see bottom panels of Figure~2). Because AR~10798 develops and maintains its anemone structure in the simulation as was observed on the Sun, we believe our model captures the most important features of the solar corona before the 2005 August 22 eruptions.

 \begin{figure*}[t]
\begin{minipage}[]{1.0\linewidth}
\begin{center}
{\includegraphics*[width=7cm]{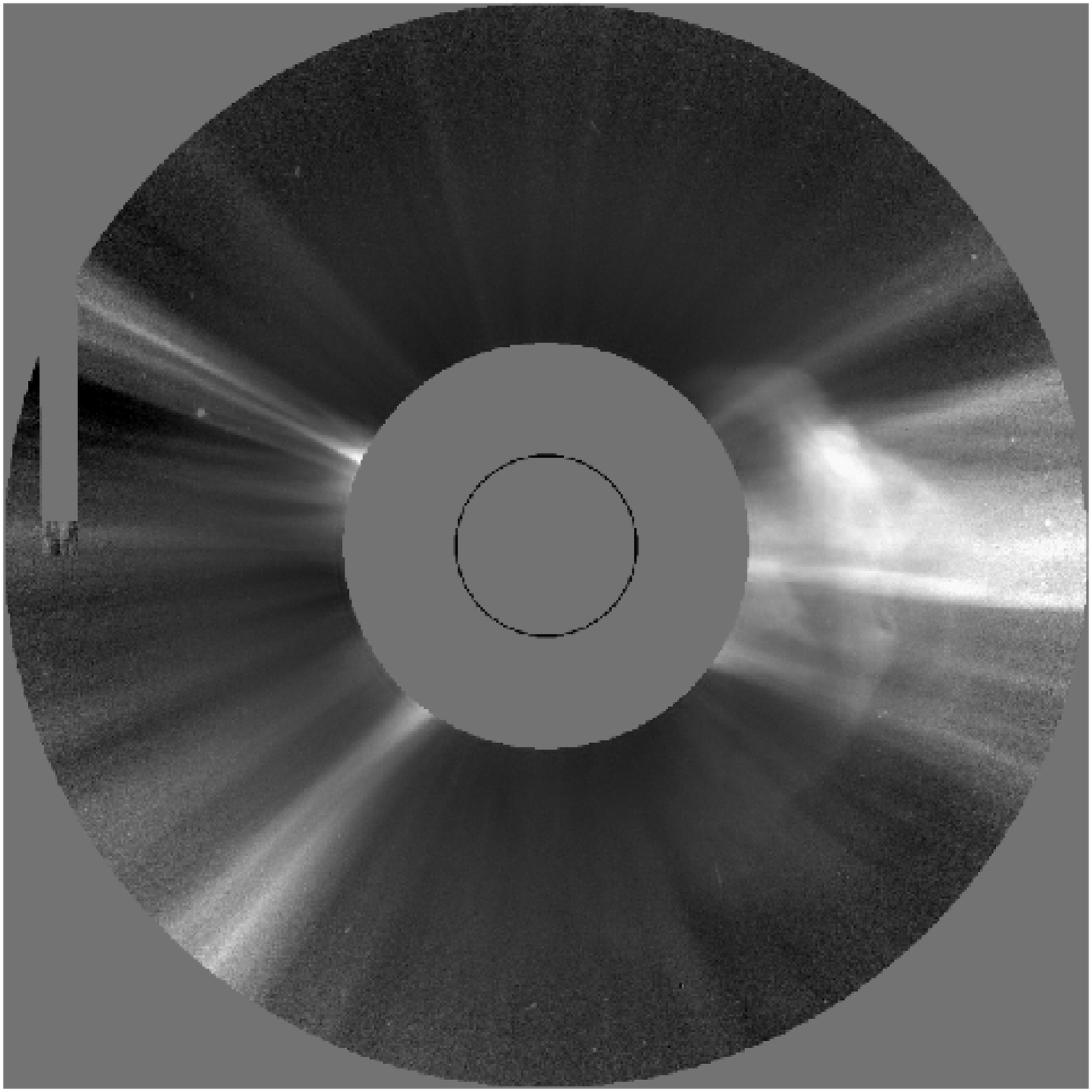}}
{\includegraphics*[width=7cm]{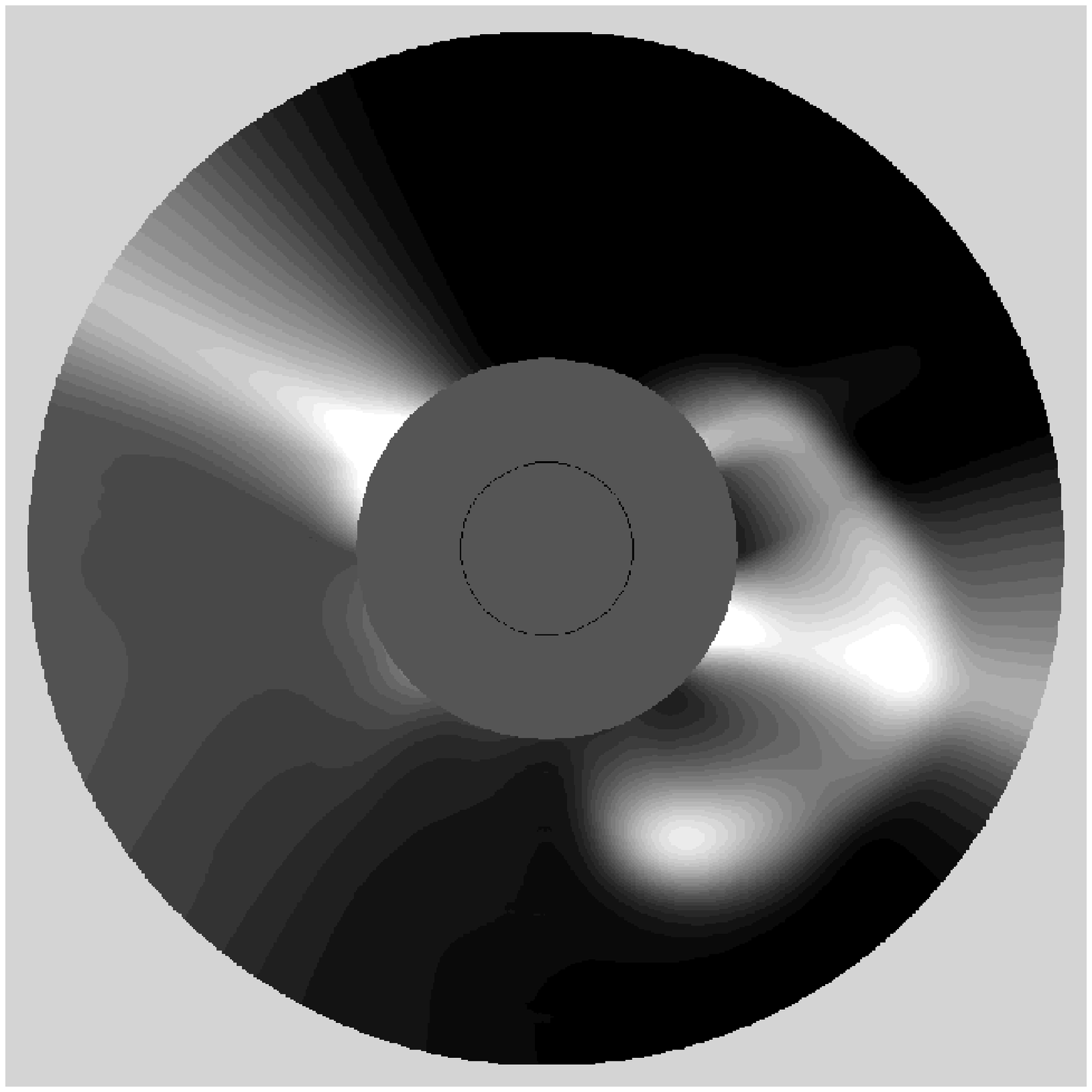}}\\
\end{center}
\end{minipage}\hfill
\caption{Line-of-sight image of the CME from LASCO/C2. The left panel is a real image 52 minutes after the start of the HXR flare and processed with the  normalized radial gradient filter \citep[NRGF, see:][]{Morgan:2006}. The right panel is a simulated image at time $t=45$ minutes after the superposition of the flux rope onto the solar surface. A simulated 27-day minimum image is subtracted and the signal further scaled with distance.}
\end{figure*}

\subsection{Pre-event Comparison of the Simulated Corona with EUV Images}
One of the advantages of using the LC model is the ability to simulate extreme ultraviolet (EUV) images following the procedure described in \citet{Downs:2010}. It allows us to validate our steady-state model by comparing synthetic images with real ones prior to the event. The EUV signal depends on the density and temperature of the plasma and it is greatly influenced by the magnetic structure of the lower corona. Therefore, this type of comparisons validates not only the plasma properties of our simulated corona but also its detailed magnetic structure. We show a comparison of an 195\AA~and 171\AA~images from the EUV Imaging Telescope \citep[EIT, see:][]{Delaboudiniere:1995} onboard {\it SOHO} 18 hours before the eruption with synthetic images for the same filters from our simulation in Figure~3. Synthetic and real images are plotted using the same scale. The 195\AA~filter response function peaks around 1.4~MK, while the 171\AA~peaks around 1~MK, illustrating different heights in the corona. In synthetic and real images, the three active regions are clearly visible as regions of enhanced emission (hot and dense): the large AR~10800 on the eastern side of the Sun near disk center, AR~10797 near the western limb and AR~10798 near W50. In addition to the northern polar coronal hole, there are a number of equatorial coronal holes, including two on the south-western side of ARs 10798 and 10800. The southern polar coronal hole is almost absent in the simulated and real images. Overall, there is good agreement between synthetic and real images, which gives us confidence that our model of the the corona is a relatively realistic representation of the actual corona at the time of the 2005 August 22 eruption. The most important features for our study that the model reproduces are the appearance of AR~10798 and the presence and aspect of open field regions (dark) around it. The main difference between simulated and real images is the eastern limb of the Sun where the modeled emission is too weak as compared to the real one. This is relatively unimportant because these regions are far from the source region of the CME (more than 120$^\circ$ separation) and are not involved in the eruption process.

 \begin{figure*}[t]
\begin{minipage}[]{1.0\linewidth}
\begin{center}
{\includegraphics*[width=5.cm]{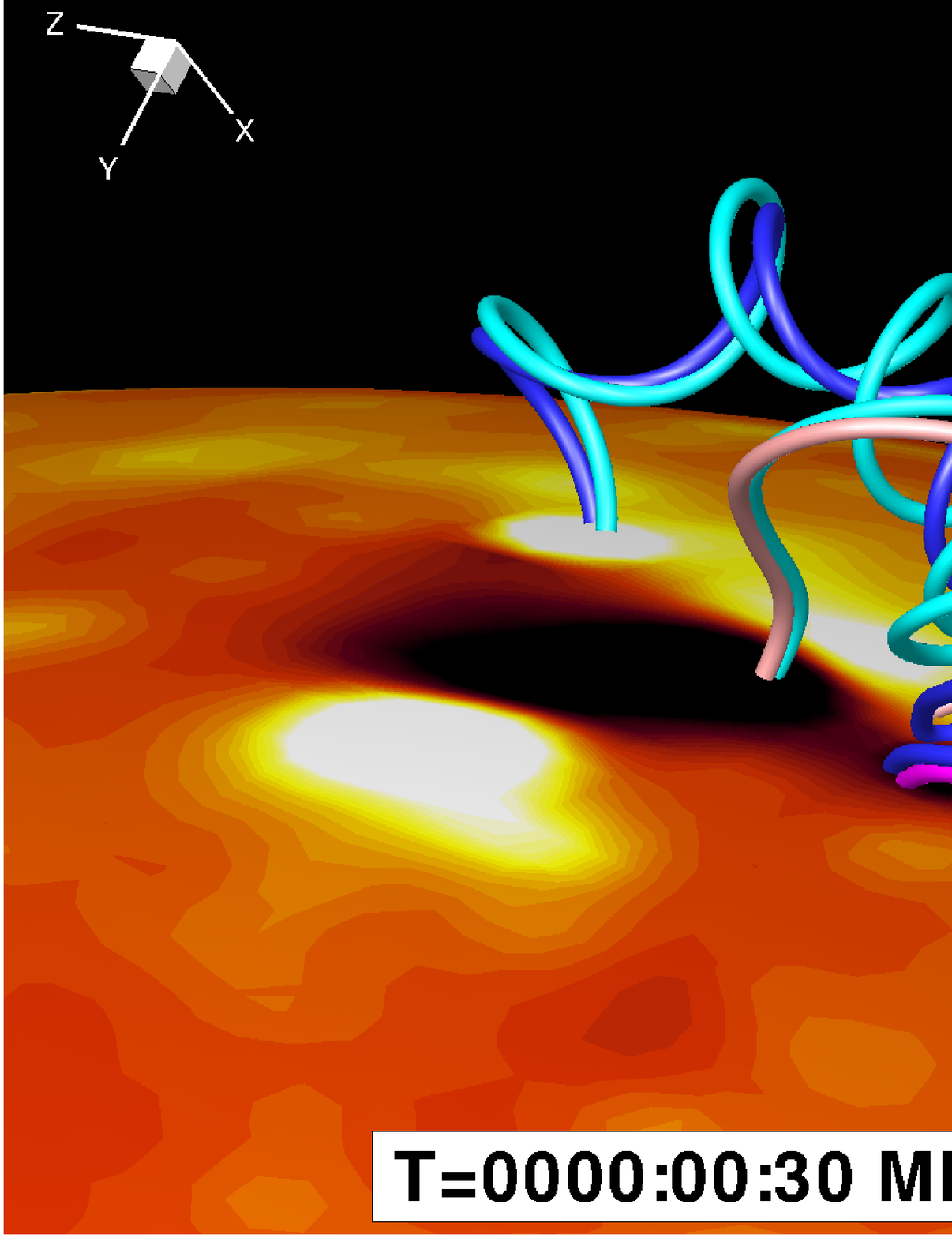}}
{\includegraphics*[width=5.cm]{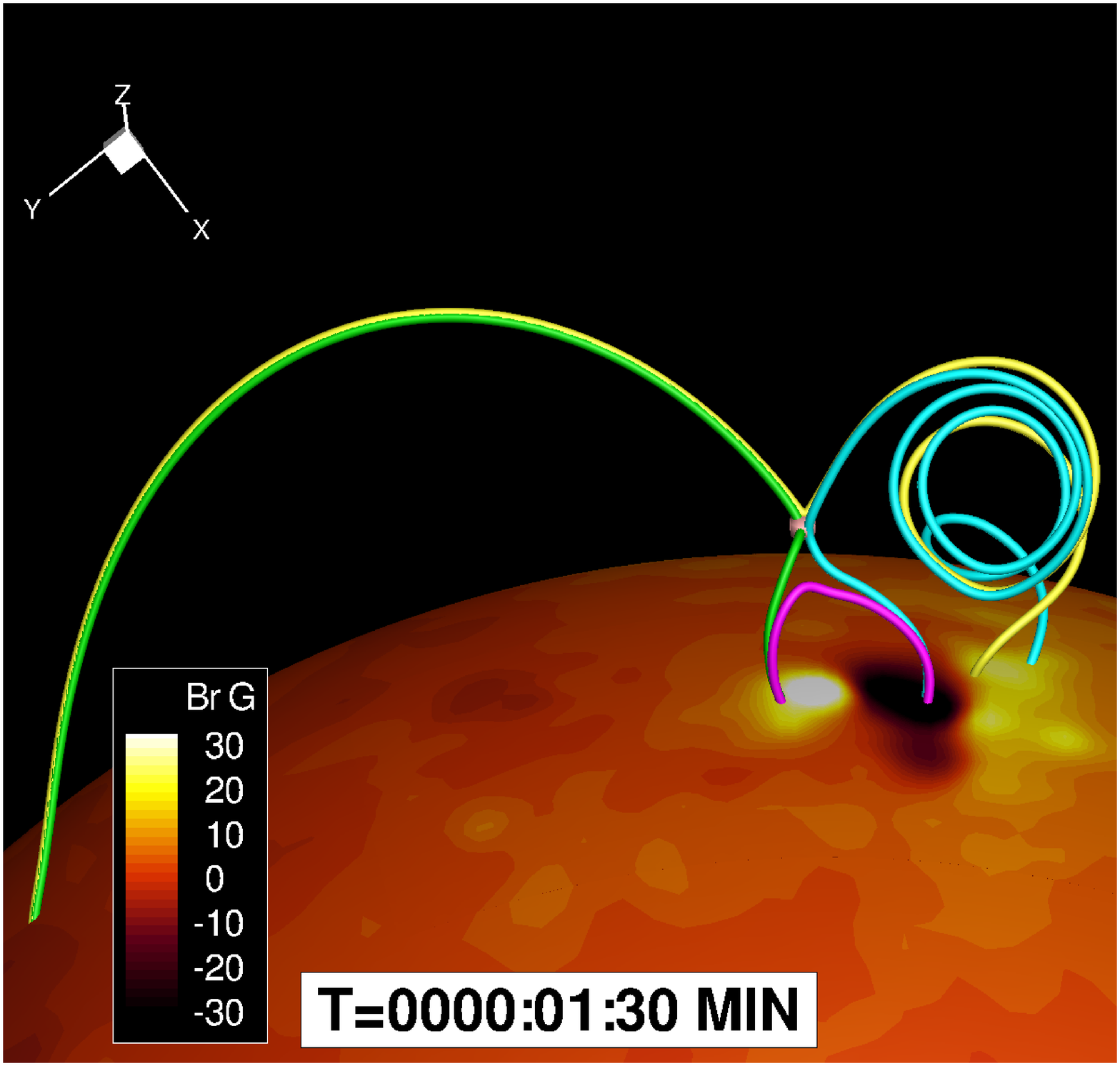}}\\
{\includegraphics*[width=5.cm]{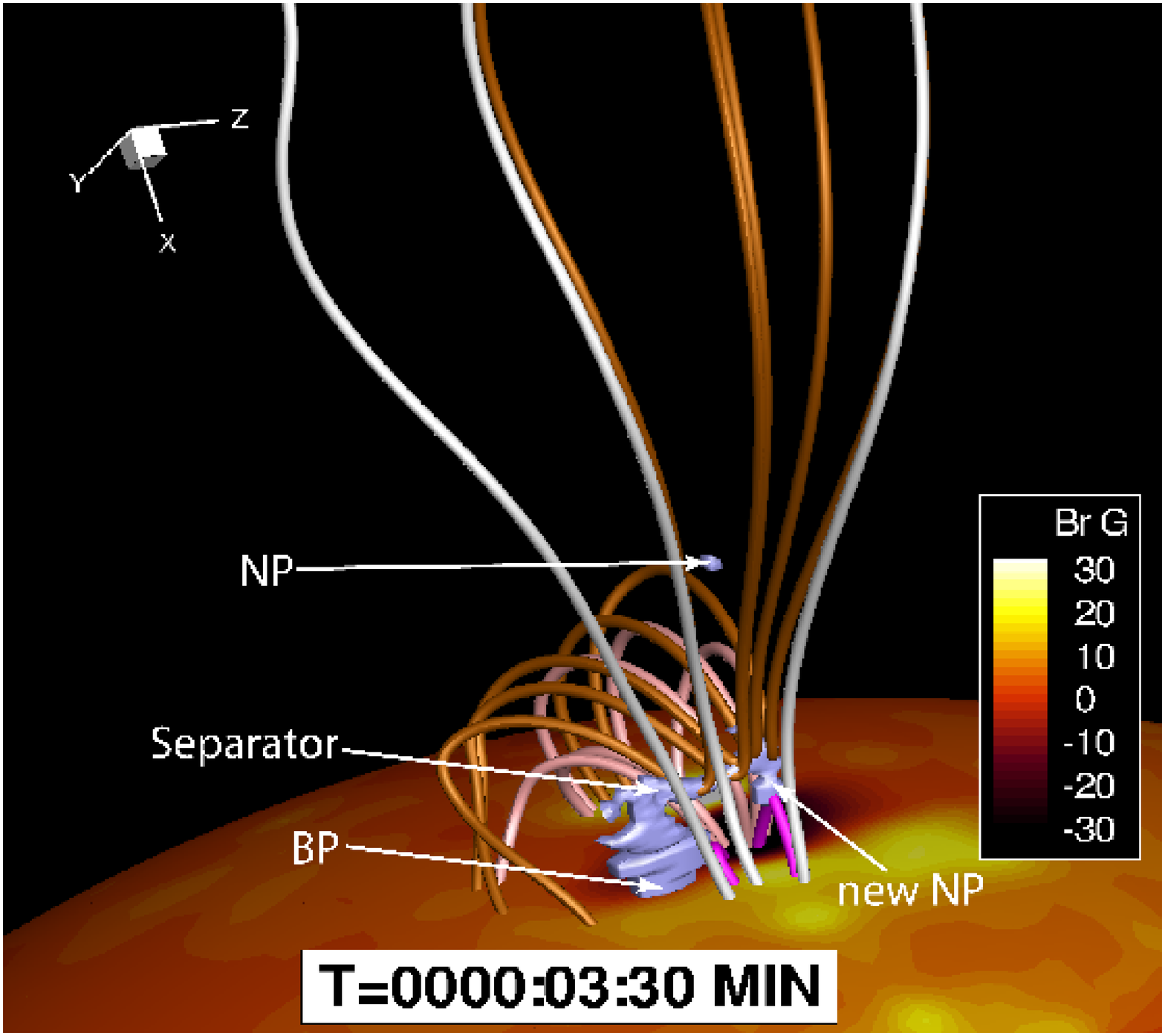}}
{\includegraphics*[width=5.cm]{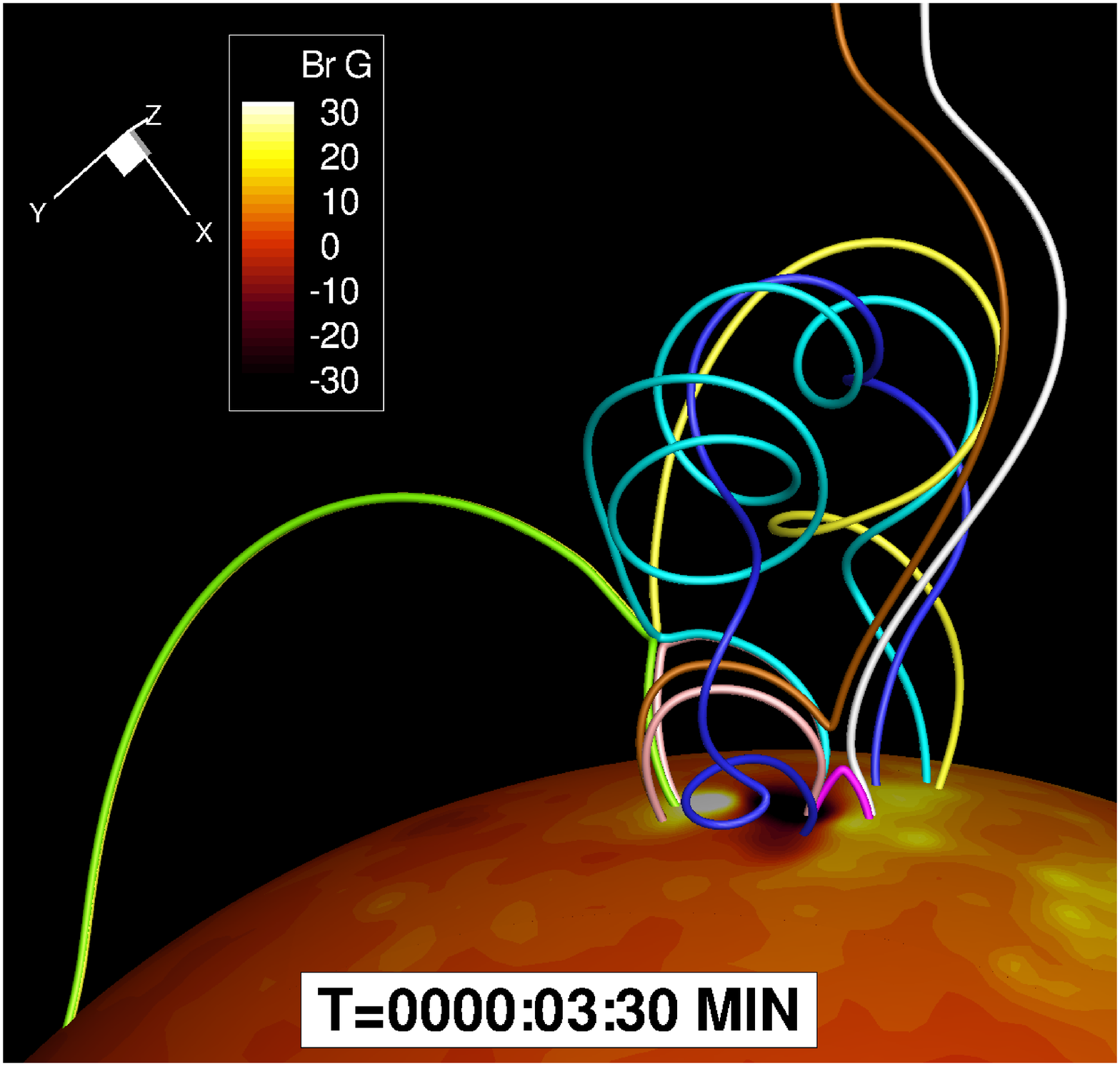}}\\
{\includegraphics*[width=5.cm]{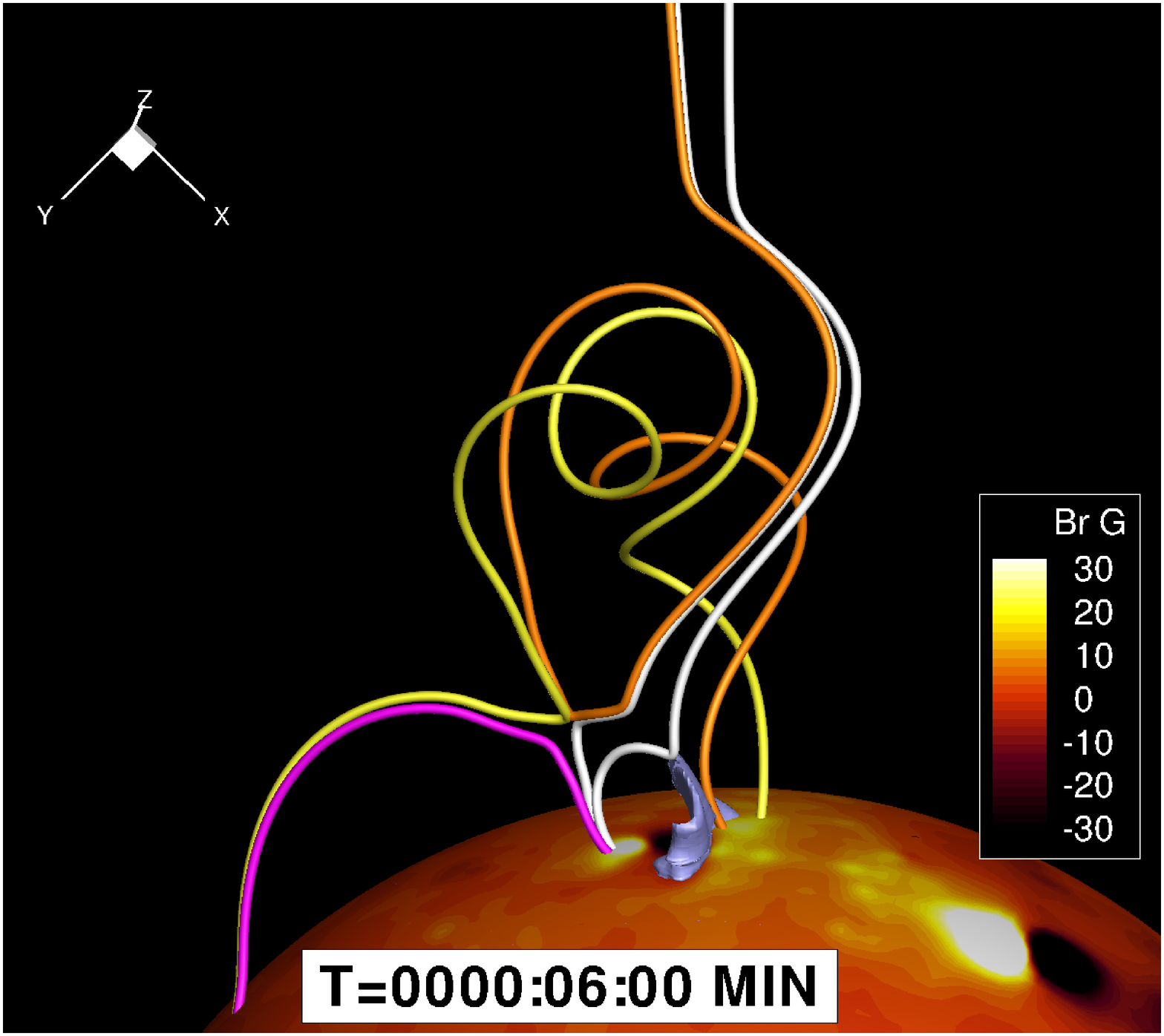}}
{\includegraphics*[width=5.cm]{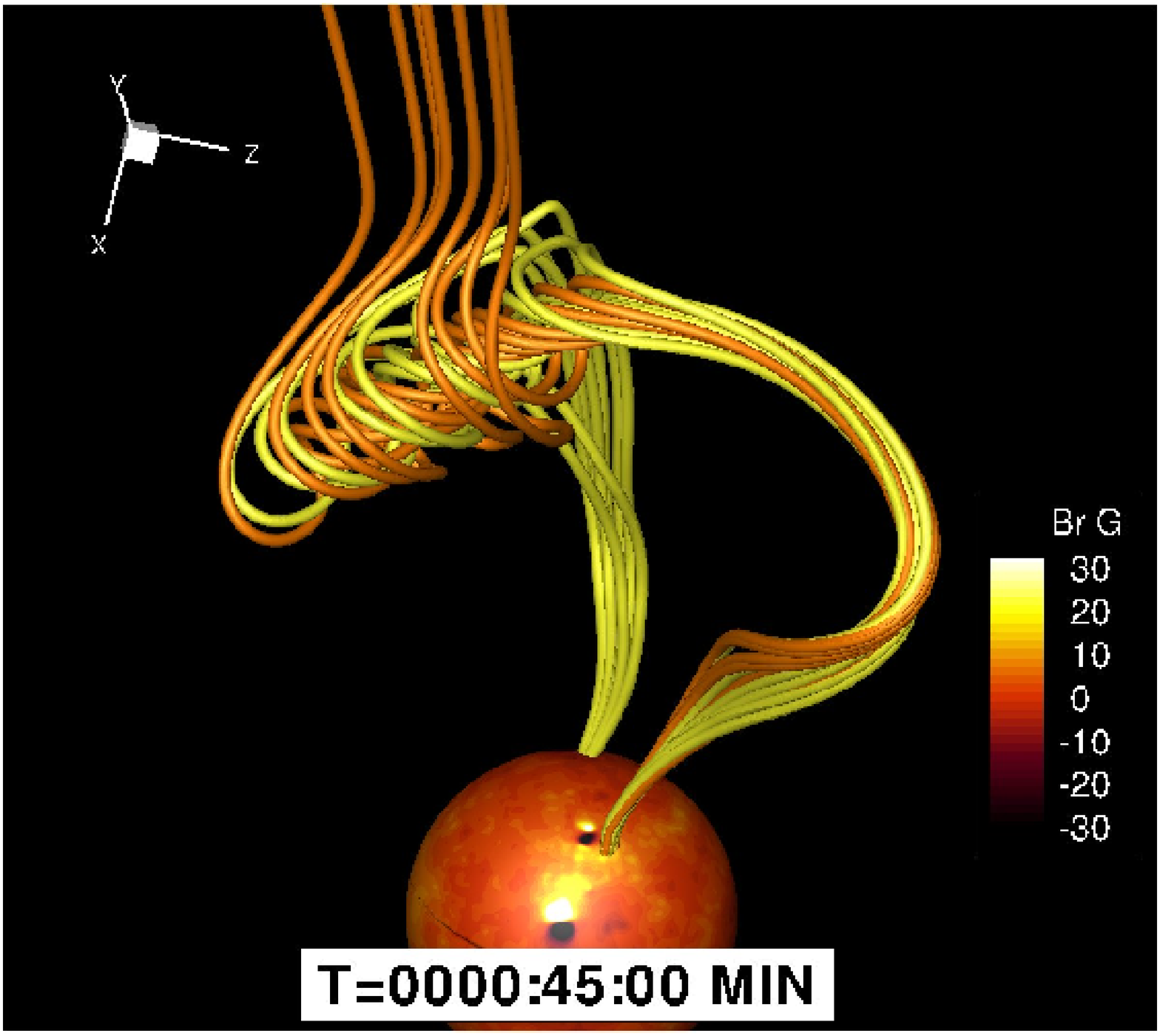}}
\end{center}
\end{minipage}\hfill
\caption{Reconnection process of the flux rope at time 0.5, 1.5, 3.5 (twice), 6 and 45 minutes after the superposition of the flux rope onto the solar corona. In all panels, the sphere is at 1.01~R$_\odot$ and is color-coded with the radial magnetic field strength. Dark blue, green, pink and white field lines show the initial flux rope magnetic field, the streamer magnetic field, the anemone magnetic field and open magnetic field, respectively. Light blue field lines are the result of the first step of the first phase of the reconnection process. Yellow and orange magnetic field lines are the result of the 2-phase reconnection process. Yellow field lines are closed and connect the positive polarity spot of the AR to the streamer belt. Orange field lines are open. Fushia field lines are post-flare loops. In the middle panels, brown field lines are newly open. In the bottom right panel, there is an equal number of open and closed twisted field lines. See online movies for a different view.}
\end{figure*}

\section{CME EVOLUTION}\label{sec:TD}

\subsection{Early Dynamics of the CME}

As soon as we superpose the flux rope onto the steady-state coronal magnetic field, it erupts due to force imbalance. As in other simulations with the same CME model, the exact kinematics of the CME early on in the corona are not realistic as the CME reaches its maximum speed ($\sim 1500$~km~s$^{-1}$) about 1.5 minutes after the superposition of the flux rope. However, the simulated CME kinematics past 3~R$_\odot$ (after 15 minutes) are in good agreement with the height-time profile as observed in LASCO/C2. Additionally, the CME speed after 1 hour is about 1200~km~s$^{-1}$, similar to what is measured with LASCO ($\sim 1250$~km~s$^{-1}$). It confirms that the total energy of the simulated CME is comparable to that of the real CME. The model does not intend to capture the slow rise phase before the loss of equilibrium nor the acceleration phase. This is why when comparing synthetic and real images, it is best not to use the onset time of the flare as the starting time of the numerical simulation but a later time when the CME has already significantly accelerated.  Previous studies have found that the CME acceleration happens during the X-Ray rise phase \citep[]{Ohyama:1998, Forbes:2000, Temmer:2008}. We use GOES-12 \citep[]{Hill:2005}  and RHESSI \citep[]{Lin:2002} data to investigate the flare time in soft X-ray (SXR) and hard X-ray (HXR), respectively. For the ejection, the flare onset in SXR was 00:44UT, the HXR flare started at 01:02UT and the flare peaked at 01:22UT. We believe the onset of the HXR flare is the best time to use for the start time of our simulation since the CME was already observed by LASCO at 4~R$_\odot$ 10 minutes after the flare peak. 

Figure~4 shows a line-of-sight image of the CME observed by LASCO/C2 and processed with the method of \citet{Morgan:2006} and a synthetic line-of-sight image from our simulation. The LASCO image is taken at 01:54 UT, 52 minutes after the onset of the HXR flare, while the synthetic image is made 45 minutes after the superposition of the flux rope onto the solar surface. The synthetic image is processed as explained in \citet{Lugaz:2009b} using a synthetic 27-day minimum image created from the steady-state simulation. The main difference between the synthetic and real images is the latitudinal direction of propagation of the CME.  In the synthetic image, the fastest moving part of the CME is at a PA of 210$^\circ$ while it is about 220--225$^\circ$ for the real image (this region is dimmer than the main part of the CME which is closer to PA 270). Because the CME appears as a halo, it is not straight-forward to determine its central PA but the synthetic image shows a CME propagating about 15--20$^\circ$ more towards the south as what appears in the real image. The active region was at the Carrington latitude South 11 (S11) when the ejection occurred and in our simulation, the CME propagates without a north-south deflection, i.e. along a latitude of S11. As noted in \citet{Asai:2009}, only the northern part of the filament erupted during the flare/ejection studied here, while the southern part erupted during the second CME on August 22. It is therefore likely that the real CME propagated more towards the north than what our simulation shows since only part of the filament erupted. With our relatively simple model to start the eruption, it is not yet possible to study partial filament eruptions. 
 
As it rises and expands, the flux rope interacts and reconnects with the adjacent magnetic flux systems. Because this is a numerical study with finite resolution and based on numercial diffusion, the timing of the reconnection might not be realistic in our simulation. However, we believe that the reconnection process itself reflects what occurred on the Sun. It is because: 1) our steady-state corona is a good representation of the pre-event Sun (see previous section),  2) the CME was due to a filament, which was originally at approximatively the same position with a similar length, and with the same orientation as our flux rope, and, 3) the CME kinematics are in good agreement with what was observed by LASCO, meaning that the total energy in the flux rope is similar to that of the CME. In the section below, we  analyze the interaction of the flux rope with the corona and in section \ref{sec:dimmings}, we present some observational consequences of the reconnection process, which further validate the discussed scenario. 

\subsection{Reconnection of the Negative Footpoint}\label{sec:reconnection}

Only a few recent numerical works have focused on the interaction of a CME with a realistic magnetic structure obtained from magnetogram measurements \citep[]{Roussev:2007, Lugaz:2010a, Cohen:2010, Evans:2011}. Other simulations have usually relied on an ideal representation of the corona with a single dipolar or quadrupolar active region in a dipolar Sun \citep[as in][for example]{Manchester:2003, Lynch:2009, Jacobs:2009, Shiota:2010}. Here, the interaction of the flux rope with the background magnetic field is different from both these ideal cases and the realistic cases previously studied  because of the anemone nature of the active region. The topology is different from the solar minimum cases studied before with an isolated AR not surrounded by equatorial coronal holes \citep[]{Cohen:2010, Evans:2011} and also different from the two solar maximum case studies of \citet{Roussev:2007} and \citet{Lugaz:2010a} with complex connectivity between multiple ARs. Recently, \citet{Titov:2008} studied in great detail the magnetic topology of the corona before and during the evolution phase leading to the 1997 May 12 CME. Their study was also for a relatively simple magnetic topology without influence from equatorial coronal holes.

 \begin{figure*}[t]
\begin{minipage}[]{1.0\linewidth}
\begin{center}
{\includegraphics*[width=6.7cm]{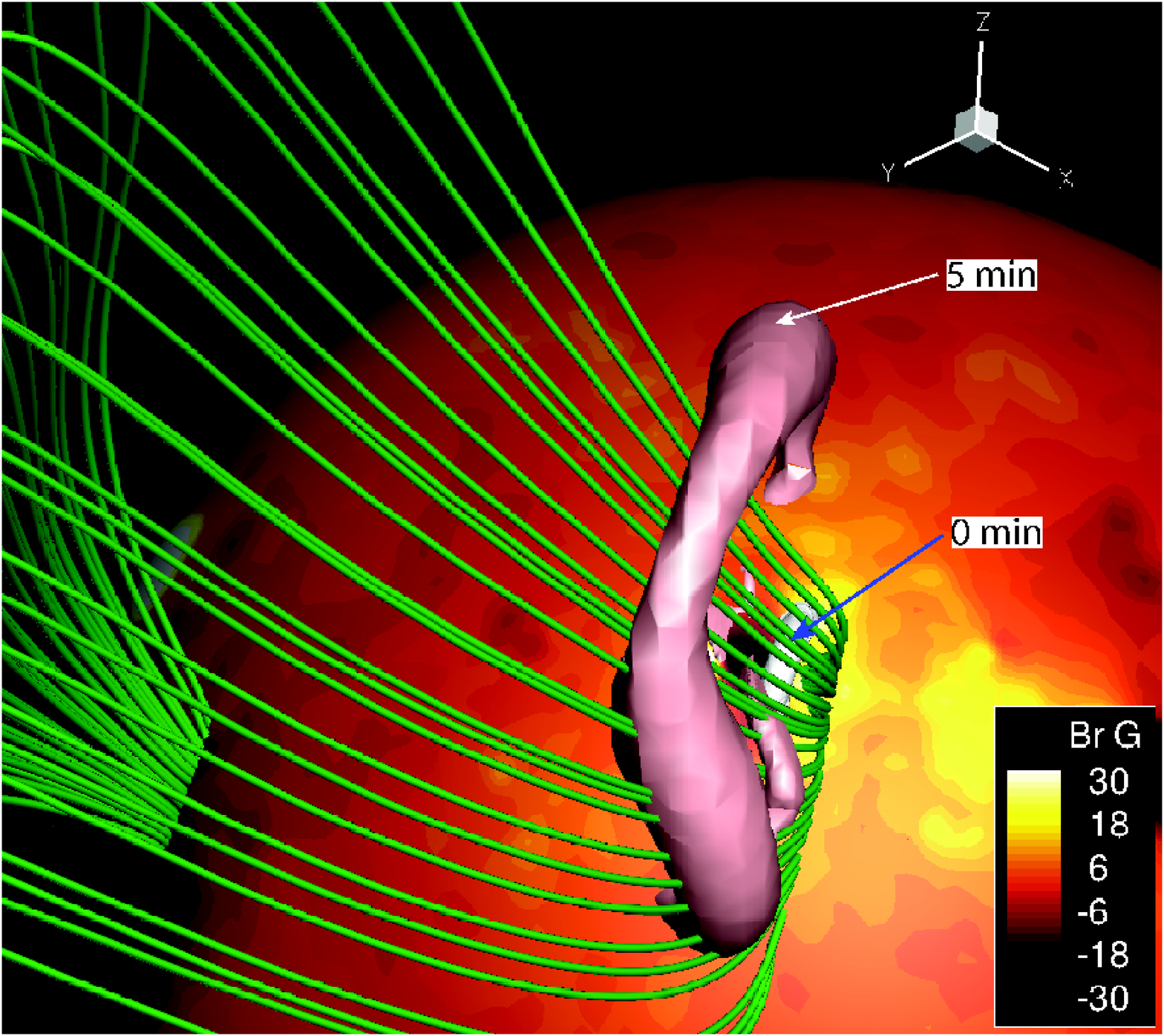}}
{\includegraphics*[width=6.7cm]{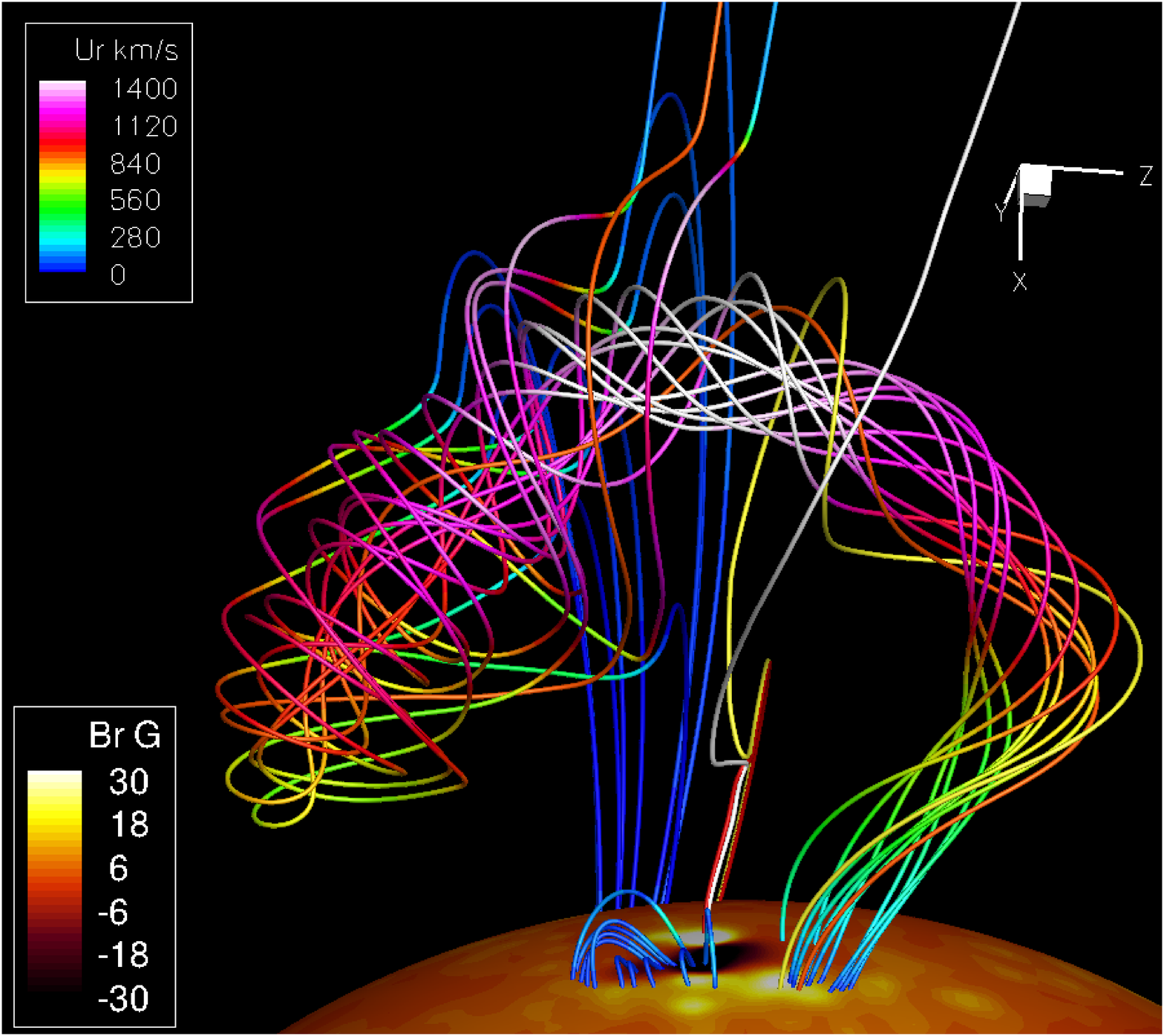}}\\
{\includegraphics*[width=6.7cm]{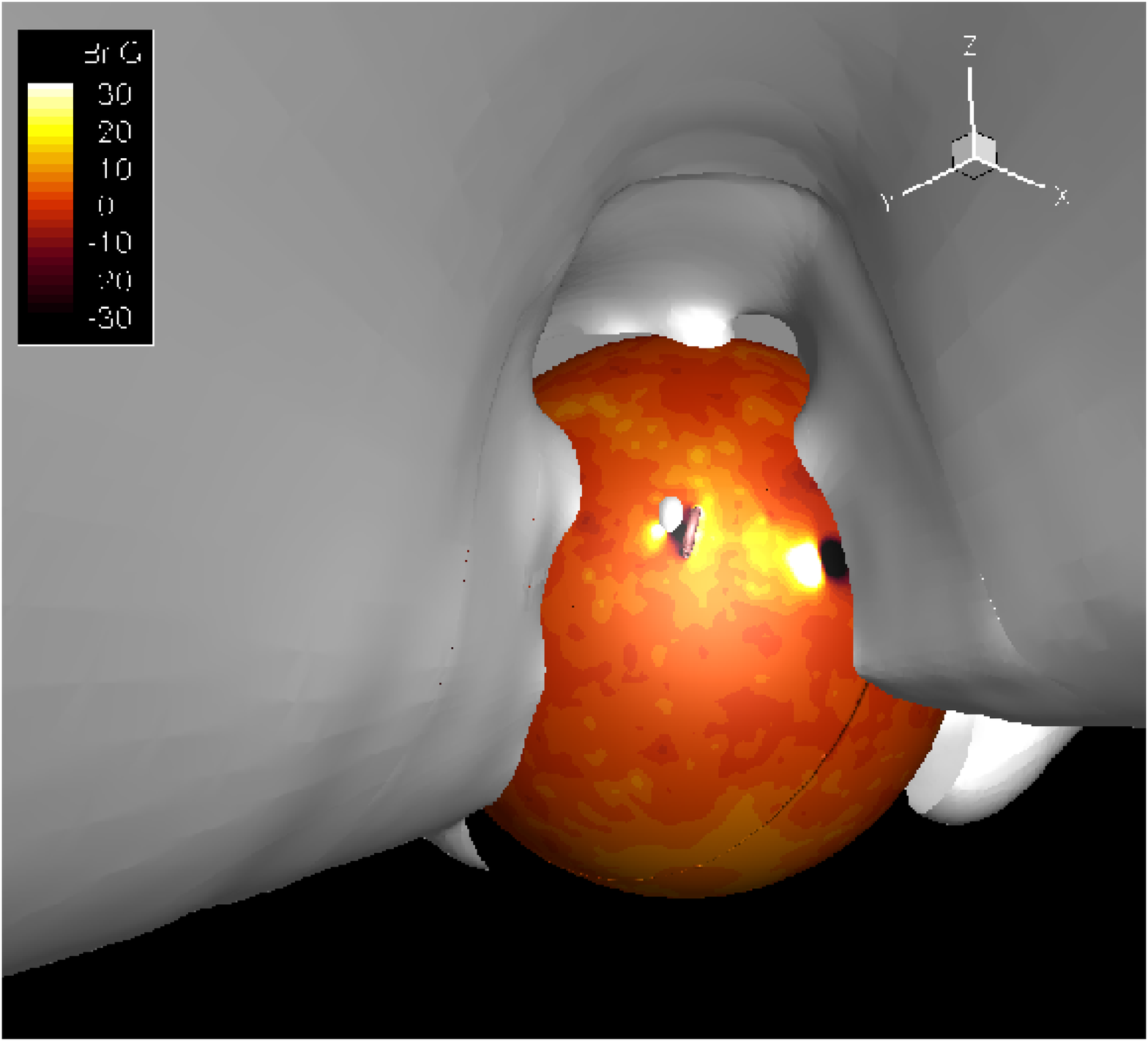}}
{\includegraphics*[width=6.7cm]{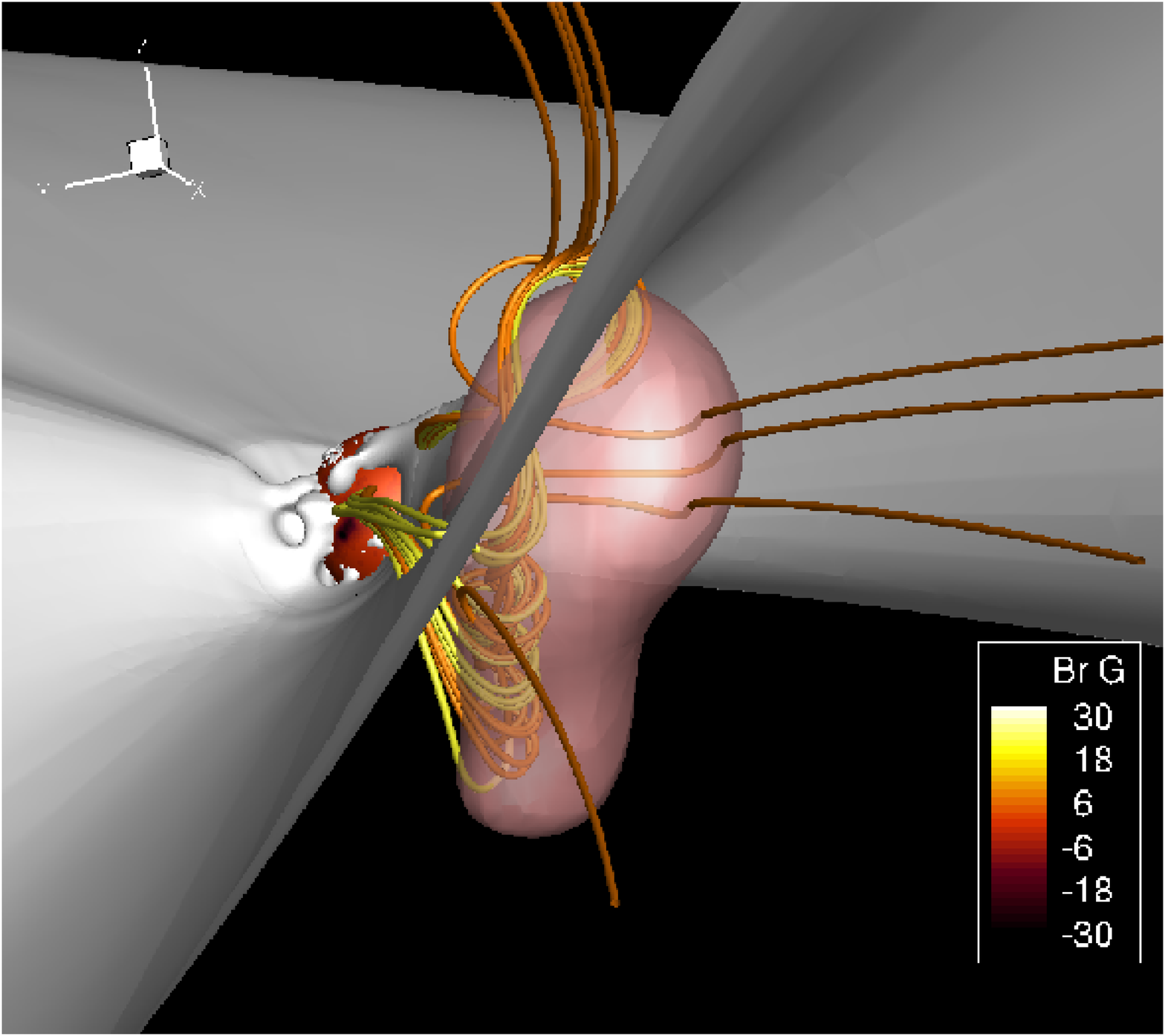}}
\end{center}
\end{minipage}\hfill
\caption{{\it Top Left Panel}: The pink surface is an isosurface of current density at time $t=5$~minutes showing the deformation of the flux rope as compared to the initial flux tube shown as a white isosurface of current density. The green field lines belong to the streamer belt at time $t=0$. {\it Top Right Panel}: Magnetic field lines color-coded with the radial velocity 6 minutes after the superposition of the flux rope. This panel illustrates the state of the flux rope after the first phase of reconnection. {\it Bottom Left Panel}: Initial orientation of the flux rope (pink isosurface) with respect to the current sheet (white isosurface). {\it Bottom Right Panel}: CME after 1 hour.  The pink transparent surface is an isosurface of velocity equal to 800~km~s$^{-1}$ illustrating the position and orientation of the CME. The white surface is the current sheet (same as left panel). It illustrates how the CME is not aligned with the HCS as it leaves the corona. For the left panels, the view is approximatively from the ecliptic plane above AR~10798. For the top right panel, the view is from the western limb looking at AR~10798. For the bottom right panel, the view is approximatively from Earth. }
\end{figure*}

The evolution of the CME is altered by the anemone nature of the active region, i.e. by the presence of unipolar magnetic field around it. We find that the positive footpoint of the flux rope does not reconnect away from the active region since there is a very limited negative polarity magnetic flux around it. As a consequence, the flux rope remains line-tied at its positive footpoint. In contrast, the negative footpoint reconnects quickly and extensively with the positive magnetic flux, partly from the positive spot of AR 10798 and partly from the neighboring coronal hole.
The reconnection is a 2-phase process as illustrated in Figure~5. In the first phase (top panels), all of the twisted closed field lines of the flux rope (dark blue) reconnect with the streamer magnetic field (green). It results in erupting field lines connecting the positive footpoint of the flux rope with the streamer, shown in yellow and post-flare loops in fushia. In the second phase (bottom panels), the newly formed (yellow) erupting field lines reconnect with open field lines to form (orange) open and twisted erupting field lines. Below, we describe the detailed reconnection process and the change in magnetic topology during the eruption.

\subsection{Details of the Reconnection Process and Changes in the Magnetic Topology}

The reconnection involves two phases: the first one, when the negative footpoint of the flux rope reconnects outside the fan surface (top panels of Figure~5) and the second one, when some of the erupting field lines open up (bottom panels of Figure~5). The detailed reconnection process of this first phase goes as follows: first, the (dark blue) field lines of the flux rope reconnect with overlying anemone field (pink) to form twisted (light blue) field lines connecting the positive foopoint of the flux rope with the negative main spot of the active region (top left panel of Figure~5). This reconnection happens at the bald patch. 
In a second step, the newly formed field lines (light blue) reconnect with closed (green) field lines outside of the fan surface to form the (yellow) field lines connecting the positive footpoint of the flux rope with the streamer (top right panel). This reconnection occurs at the null point. 
The two steps of the first phase are nearly simultaneous and overlying closed field lines from the streamer belt flux system (green field lines) reconnect away.
 
During the reconnection process, a null point forms below the flux rope and there is a separator linking the bald patch and this null point. This separator is similar to the bald patch-bald patch separator or the quasi-separatrix layer discussed for example in \citet{Titov:1999} and \citet{Aulanier:2010} and it is shown in metallic blue in the middle left panel of Figure~5. A current sheet forms in association with the separator and it is shown as a purple surface in the bottom left panel of Figure~5. The current sheet takes a sigmoidal shape, which is very common from flux emergence and shearing motions \citep[]{Manchester:2004c,Titov:2008, Aulanier:2010} but it can also be the result of reconnection with adjacent flux systems and due to the pre-event topology as is found here.

During the first phase, which lasts about 5--10 minutes, the flux rope expands to a height of 1~R$_\odot$ above the solar surface. Due to the CME expansion, the null point is pushed eastward along the outer spine by about 3$^\circ$ and its height increases by about 0.04~R$_\odot$. At the end of this phase, the flux rope remains composed of closed field lines but it appears totally disconnected from its original negative footpoint (see top right panel of Figure~6). 

Open field lines (white) from the leading positive spot reconnect at the separator below the flux rope with anemone field lines to form new open field lines (shown in brown) passing at proximity of the null point. Separator reconnection is described in \citet{Parnell:2004} for example. This is illustrated in the middle left panel of Figure~5. A similar type of reconnection following the eruption of a flux rope has been previously discussed in \citet{MacKay:2006}. 
Due to the reconnection of the anemone flux system and the streamer belt and due to the formation of the separator, the magnetic topology changes quite drastically from the pre-event topology. The null point below the flux rope separates the anemone flux system and open field lines originating from the leading (white) and trailing (brown) positive spots (middle left panel of Figure~5). The outer spine of this system is open. Such an opening of a similar topology during a time-dependent process was shown in \citet{Edmondson:2010}. The eastern null point separates the erupting (yellow) magnetic field and the closed field originating from the positive trailing spot. The outer spine connects the positive trailing spot and the positive footpoint of the flux rope through the flux rope. The full magnetic topology with all the different type of field lines discussed here is shown in the middle right panel of Figure~5. The presence of two null points separating different but overlapping flux systems, as shown here, is only possible in a fully 3-D simulation as is the case here. The second phase of reconnection happens when the flux rope is not connected anymore to the negative spot of AR 10798. 

 \begin{figure*}[t]
\begin{minipage}[]{1.0\linewidth}
\begin{center}
{\includegraphics*[width=7.8cm]{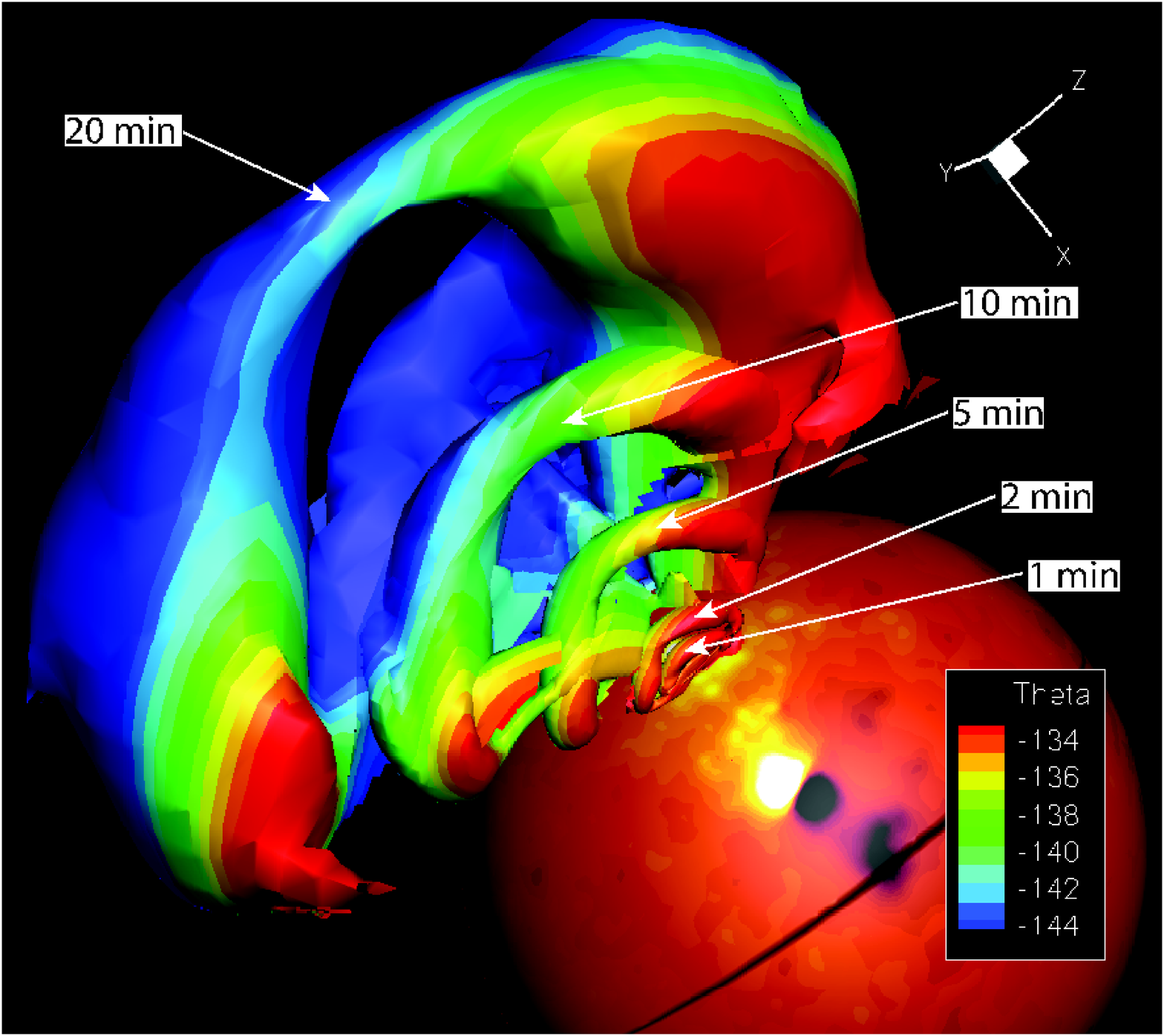}}
{\includegraphics*[width=8.2cm, height = 6.9 cm]{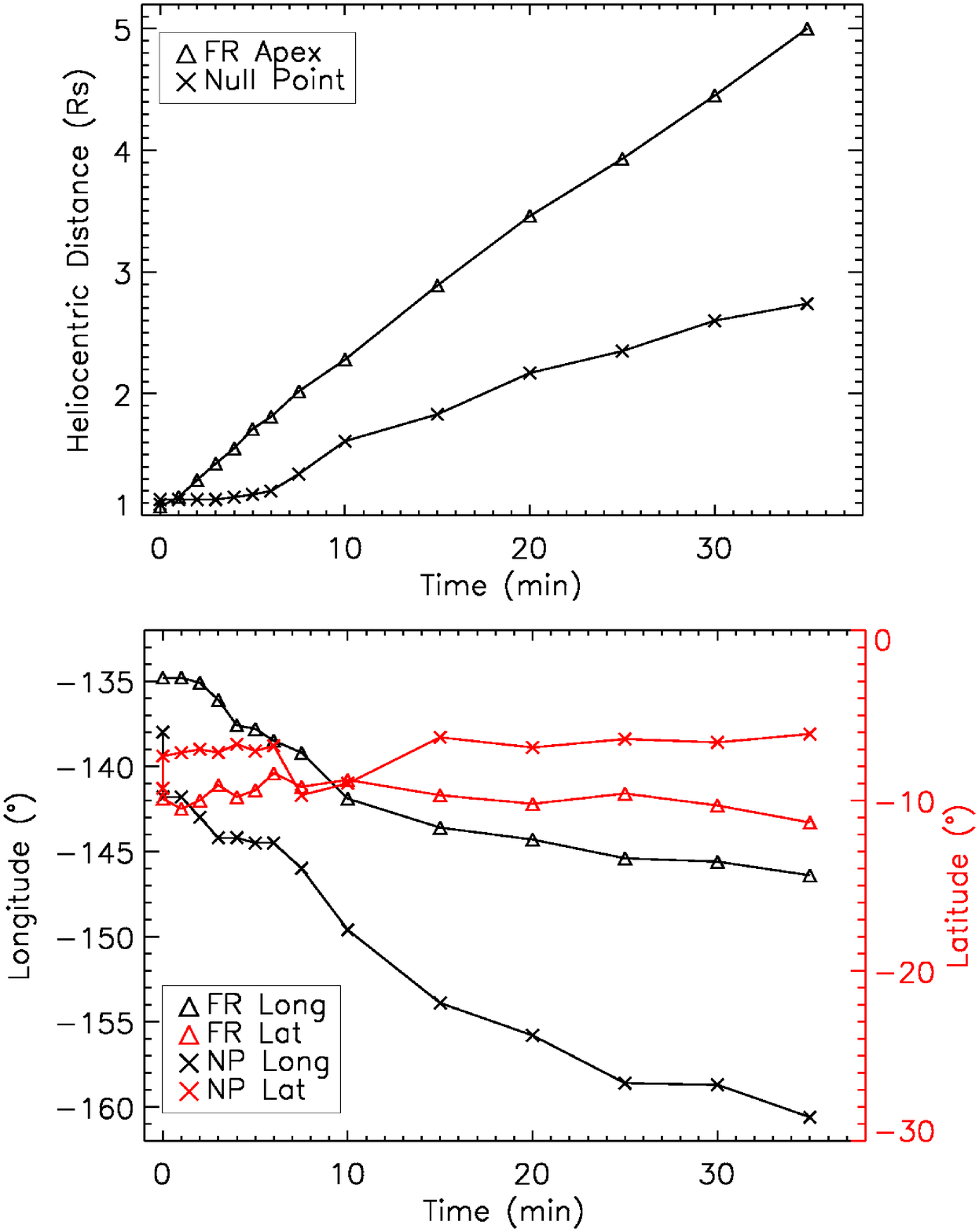}}
\end{center}
\end{minipage}\hfill
\caption{{\it Left panel}: 5 views of the flux rope at time 1, 2, 5, 10 and 20 minutes after the superposition of the flux rope onto the solar surface. Each surface is an isosurface of current density and they are color-coded with the longitude. As the color contour goes from dark red to dark blue, the longitude varies from 0 to 10$^\circ$ east of the location of the active region, illustrating the CME deflection. The initial longitude of the flux rope in the coordinate system of the Figure is $-134.5^\circ$; after 20 minutes it is about $-143^\circ$. This panel also illustrates how the CME is deformed but does not rotate. This view is from above the western limb looking down at AR~10798. {\it Right panel}: Evolution of the apex of the flux rope and that of the pre-existing null point (top: distance; bottom: latitude --red-- and longitude --black--).}
\end{figure*}

The second phase of the evolution starts when the negative footpoint of the flux rope has fully reconnected outside of the fan surface (light and dark blue field lines no longer exist)s. Then, the flux rope has raised, expanded and deflected enough so that {\bf side} reconnection happens with open magnetic field lines at the null point \citep[similar to what was found in][]{Chen:2000}. The reconnection of closed erupting field lines with open field lines from an equatorial coronal hole is a type of interchange reconnection. Here, contrary to the case described in \citet{Crooker:2002}, the open field lines are only transported over a very small distances as they reconnect with closed field lines, because the entire active region is surrounded by equatorial coronal holes. This special form of interchange reconnection is due to the anemone nature of the active region and one can expect that it happens in most eruptions from an active region inside a coronal hole. It has been previously discussed in an observational study of an eruption from an anemone active region in 2007 October 17 \citep[]{Baker:2009}. 

At the time the second phase of reconnection starts, about six minutes after the CME initiation, the apex of the flux rope is at 1~R$_\odot$ above the solar surface and the null point is only at a height of 0.2~R$_\odot$. In this phase (bottom panels of Figure~5), the yellow erupting field lines connecting the positive footpoint of the flux rope to the streamer reconnect with white open field lines to form open, twisted field lines (shown in orange color). During this phase, the null point rises and is pushed eastward. Its evolution as well as that of the apex of the flux rope is shown in the right panels of Figure~7. Note that at all time, the null point is below and to the east side of the flux rope. An online animation shows the first ten minutes of the CME evolution with the 2-phase reconnection.

 \subsection{Mix of Closed and Open Field and Dimming Regions}\label{sec:dimmings}
 
The end result of the reconnection described in the previous subsection is typically an opening of the flux rope magnetic field but it is a relatively complex process and, importantly, some of the magnetic field lines of the flux rope remain closed until the end of our simulation (see bottom right panels of Figure~5 and 6). As seen in the bottom right panel of Figure~5, at time $t=45$~minutes, the CME is composed of a mix of closed (yellow) and open (orange) field lines. The closed field lines are highly twisted and they have, in general, one footpoint (positive polarity) near the source region of the CME and the other footpoint (negative polarity) in the quiet Sun due to reconnection with the streamer belt. The open field lines are also highly twisted because they are due to the reconnection of the magnetic field of the flux rope with that of the background originating from the coronal hole just south-west of AR 10798 (positive polarity).  This picture is similar to the sketches of \citet{Gosling:1995} with open plasmoid field line embedded inside closed twisted field lines.  
 
Figure~8 shows a base difference of the 195\AA~image 95 minutes after the start of the eruption with a pre-event image for the simulation (left) and from EIT (middle).
As highlighted in this Figure, the presence of erupted field lines with a footprint at the edge of the streamer belt near disk center results in coronal dimmings from this region in our synthetic EUV images. Theses coronal dimmings were clearly visible in EIT images on the same region of the solar surface from about 01:26 UT on August 22 to the end of August 23. Dimming regions have been analyzed as the location of the footprint of the magnetic cloud \citep[]{Webb:2000, Mandrini:2007} and/or the region where interchange reconnection happens \citep[]{Attrill:2006, Attrill:2007}. Here, we find that the dimming region corresponds to the footprint of the magnetic cloud since erupting closed field lines have their negative footprint in this region.  
We believe that the appearance of the dimming region is only possible if the first phase of the reconnection process indeed happened as described in the previous section. It is hard to understand otherwise how the filament could reconnect with a distant region when reconnection with the adjacent coronal hole should happen preferentially. 

Our simulation also shows how a flux rope originating from an active region inside a coronal hole can still contain closed field lines all the way to the outer corona (10~R$_\odot$). On 2005 August 24, ACE measured the passage of a small magnetic cloud, which has been associated with this eruption \citep[]{Asai:2009}. While the smooth rotation only lasted for about 2 hours, it implies that twisted field lines associated with one of the two eruptions on August 22 made it to 1~AU. Figure~8 also includes a base difference image about 7.5 hours after the start of the eruption to illustrate the persistence of the dimming region. The dimming region was visible until the end of August 23 and it is likely that it persisted until around 12:00UT on August 24 but the observing geometry makes it hard to assess (the dimming region had moved close to the western limb of the Sun on August 24). The fact that dimming from the location of the negative footprints of the CME persisted  until then would corroborate the measurement of a magnetic cloud at 1~AU at this time.

\section{CME DEFLECTION, DEFORMATION AND ROTATION}\label{sec:rotation}
\subsection{CME Deformation and Rotation}

 \begin{figure*}[t]
\begin{minipage}[]{1.0\linewidth}
\begin{center}
{\includegraphics*[width=5.2cm]{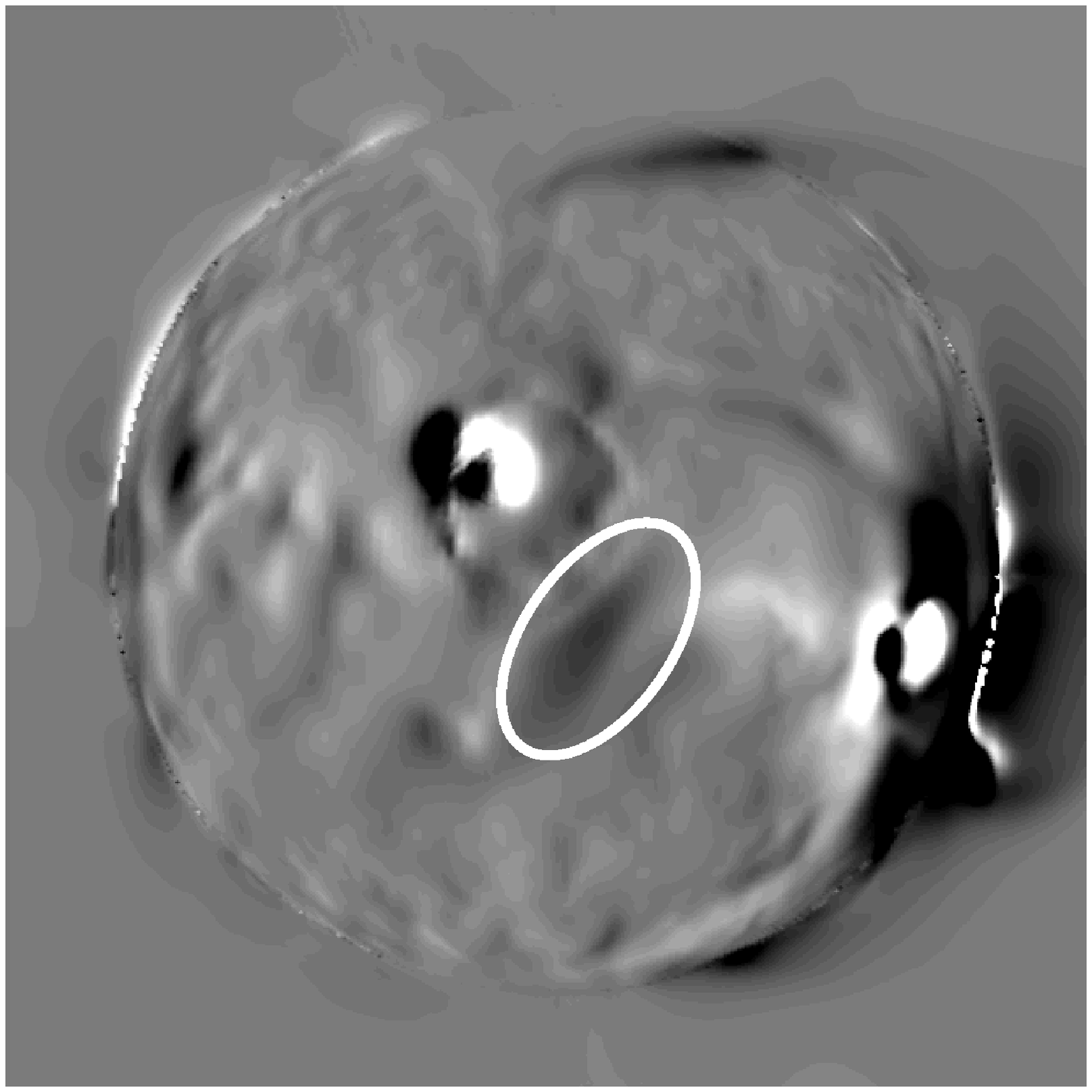}}
{\includegraphics*[width=5.2cm]{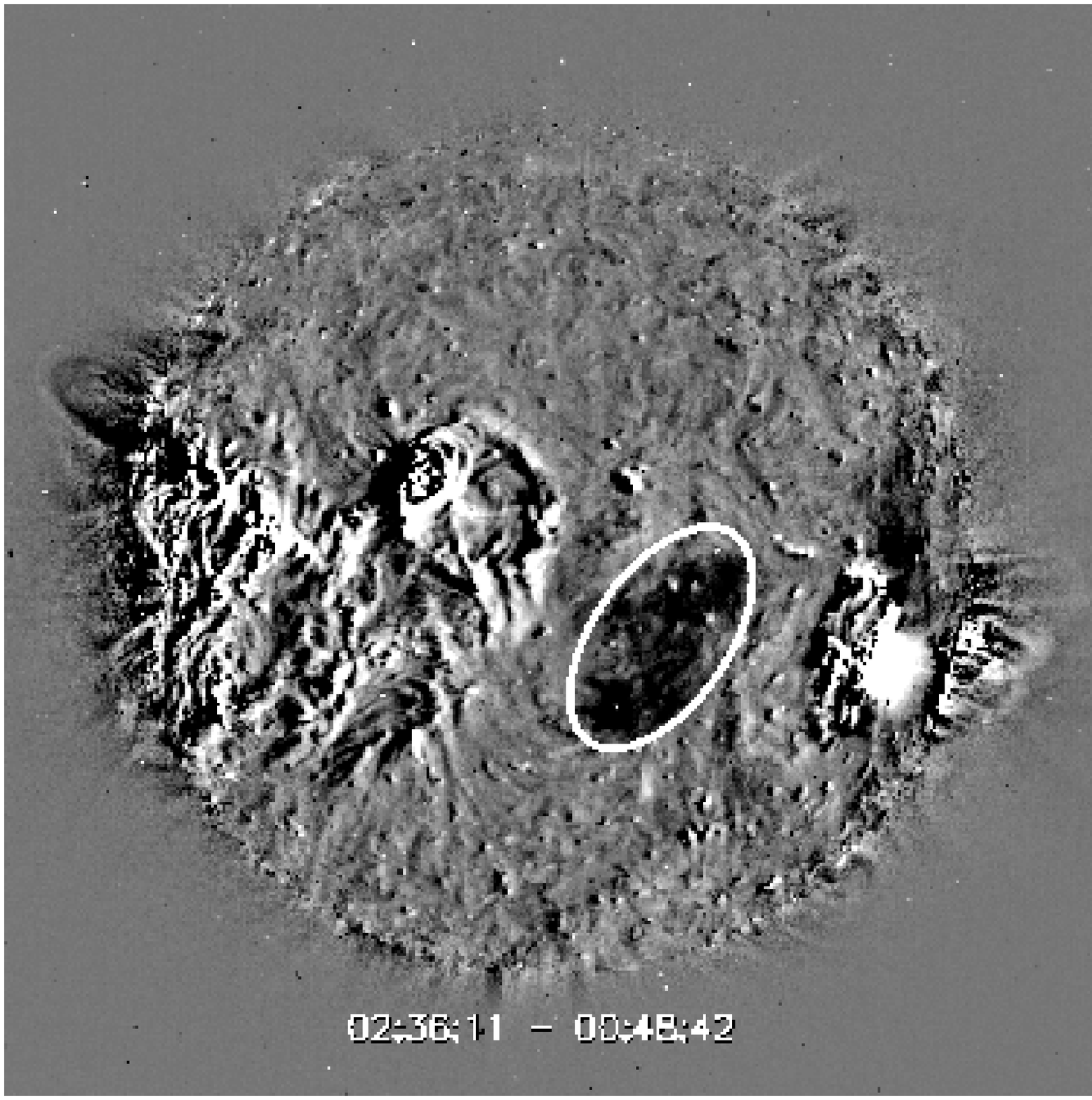}}
{\includegraphics*[width=5.2cm]{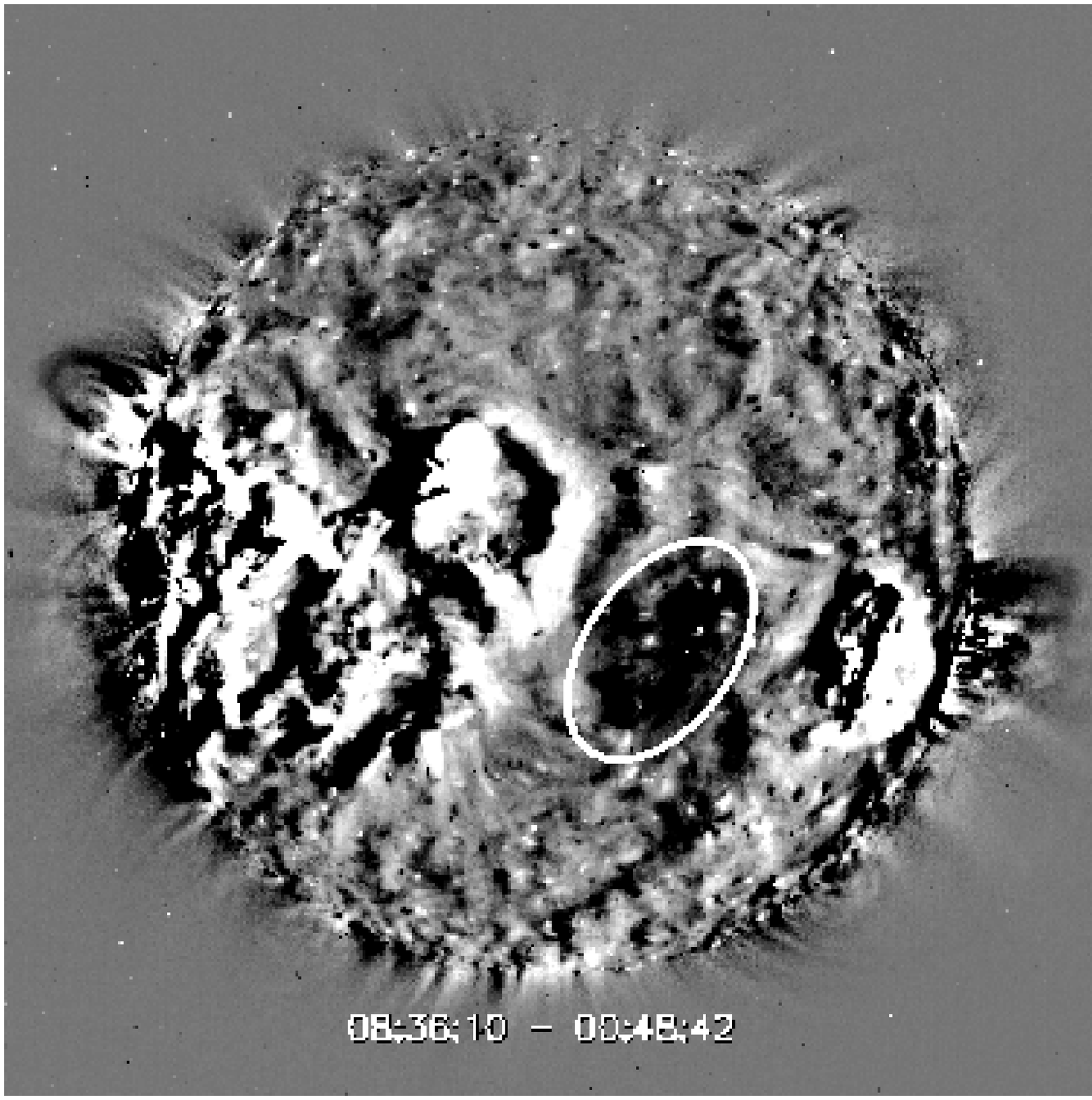}}
\end{center}
\end{minipage}\hfill
\caption{EUV 195\AA~base difference images of the CME showing the simulated corona at time $t=95$ minutes (left) and a EIT image 94 minutes after the start of the HXR flare (middle) and about 7.5 hours later (right). In all images, the dimming region corresponding to the negative footpoint of the closed erupting field lines is highlighted by a white ellipse.}
\end{figure*}

In the top right panel of Figure~6, we show the flux rope at time $t$ = 5 minutes as a pink surface as well as the initial streamer field lines (green) and the initial flux rope position (white surface) corresponding to $t$ = 0. From this Figure, it is clear that the flux rope has been deformed from a straight tube into one with a sigmoid-like shape. The reason for the deformation is the reconnection with the streamer magnetic field lines (as discussed in the previous section). The reconnection happens at the negative footpoint of the flux rope (the one on the southern part of the sunspot) and it moves the footprints of the erupting field westward to the initial position of the streamer field lines, as is clear from this Figure. The reconnection results in a rotation of the part of the flux rope which reconnects. It is similar to the rotation found in \citet{Cohen:2010} but, here, we find that it only affects part of the CME. Also, the net results is not to align the CME with the current sheet but to deform it to make part of the CME aligned with the streamer belt. Because only part of the flux rope ``rotates'', it leads to a deformation, which makes the flux tube looks similar to a sigmoid with a marked skewness. This type of skewness has long been thought to be associated with shearing motions and it is produced by many CME initiation mechanisms involving shearing motions \citep[]{Titov:2008, Lynch:2009, Amari:2010, Aulanier:2010} and flux emergence \citep[]{Manchester:2004c,Fan:2005}. Here, we find that a straight flux rope may naturally develop a skewness as it interacts with the background magnetic field at a height as low as 0.7~R$_\odot$. This finding is in addition to the sigmoidal current sheet developing under the flux rope as discussed in section 3.2.

We calculated the orientation of the central axis of the flux rope as it propagates from a distance of 1~R$_\odot$ to 8.5~R$_\odot$ (first hour) and we found no indication of rotation with an axis making an angle of -15 to -20$^\circ$ with the $-z$ axis. The bottom right panel of Figure~6 shows the CME after 1 hour as it has reached 8.5~R$_\odot$. The axis of the CME as illustrated by the magnetic field lines is found to be almost aligned with the initial axis of the flux rope. It is also illustrated in the first panel of Figure~7 with 5 views of the flux rope at different times from 1 to 20 minutes (corresponding to distances from 1 to 3.5~R$_\odot$). In this Figure, the flux rope is plotted at different times as isosurfaces of current density. The color contours show the CME longitude. The angle between a given color and the CME axis gives an indication of the CME orientation. The angle is approximatively constant over the 20 minutes covered by this Figure, illustrating the lack of rotation. 
The magnetic field measurements at 1~AU (see Figure~9 of \citet{Asai:2009}) are consistent with a right-handed flux rope with a dominant southward axial component. It is the same as the orientation of the filament on the solar surface and, therefore, implies that there was no large-scale rotation during the CME propagation. 

\citet{Lynch:2009} proposed that the rotation of a CME in the corona is related to its skewness. It would be due to the force balance required to maintain this skewness which is lost as the CME erupts. It generates a torque created by the now unbalanced Lorentz force. However, here, the skewness is not the result of shearing motions and it is not associated with force imbalance during the CME propagation. Therefore, our conclusion that the CME does not rotate is also consistent with this scenario. This result provides a caution on the association of skewness with CME rotation as it was found, here, that CME skewness can developed during the eruption, in which case no rotation is expected. 

\citet{Cohen:2010} found  a strong rotation ($\sim 90^\circ$) associated with the reconnection of a CME with the background magnetic field, yielding a CME whose axis is aligned with the heliospheric current sheet \citep[as in][]{Shiota:2010}. In our case, we found that only part of the flux rope reconnected, resulting in a deformation instead of a rotation of the flux rope axis. Although reconnection continues, it occurs primarily with open magnetic field lines, resulting in no change of the CME axis. Because there is no observational evidence that the CME rotated, it is important to understand why it is the case, when most recent simulations have had a tendency to show that CME rotation is a very frequent phenomena. Here, we find that the reconnection of one footprint of the CME results in a deformation of the CME but not a rotation and that the effect of the magnetic forces is to deflect the CME not to rotate it (see next section).

\subsection{CME Deflection}

Figure~7 illustrates the deflection of the flux rope. As shown in the right panels, the apex of the flux rope is found to deflect by about 10$^\circ$ in 35 minutes as it propagates to 5~R$_\odot$. After this time, the current inside the flux rope becomes so small that it is impossible to track the flux rope the same way as we did for the first 5~R$_\odot$. Instead, we track the position of the velocity disturbance associated with the CME. We find that the CME further deflects by about 1.5$^\circ$ from $t=35$ minutes to $t=1$ hour, at what time the CME has reached 8.3~R$_\odot$. In the absence of a velocity shear in the solar wind which could ``push'' the CME away from the coronal holes, there are two possible causes for the deflection: i) the side reconnection with the null point which happens at the east side of the CME and ii) the effect of the Lorentz force.

The flux rope is created by a southward line current and the global magnetic field above AR~10798 is of unipolar positive polarity. The Lorentz force acting on the flux rope is, therefore, almost purely eastward. The Lorentz force here is similar to that of the models of \citet{Chen:1996} and \citet{Isenberg:2007} but in these studies, it results in acceleration and rotation, respectively due to different magnetic topologies. Here, due to the largely unipolar polarity, the Lorentz force can deflect the CME. Side reconnection resulting in CME deflection can be seen in the work of \citet{Chen:2000} for example. It is not straight-forward to determine the cause of the CME deflection. However, during the first phase of the reconnection (first six minutes), the main part of the flux rope appears to be relatively unaffected by the reconnection as its negative footpoint reconnects (see top left panel of Figure 7, for example). In addition, the reconnection is a breakout type reconnection (above the flux rope) and not a side reconnection. No deflection has yet been reported for asymmetric breakout simulations \citep[e.g., see][]{Lynch:2009}. In addition, the deflection rate of the null point does not correspond to that of the apex of the flux rope, as shown in the right panel of Figure~7. During the second phase of the reconnection, the side reconnection results in an opening of the magnetic field of the flux rope and it is unclear whether it affects or not the CME direction. For these reasons, we believe that it is likely that the Lorentz force is the main contributor to the deflection of the flux rope.

The Lorentz force is expected to decrease as the CME rises since both the magnetic field strength and the current decrease with height. We find that the deflection rate decreases with height and the CME only deflects by about 2$^\circ$ in the second 30 minutes of its propagation. It is consistent with the Lorentz force being the cause of the deflection. The deflection of the actual filament at the Sun could be greater than what is found in the simulation because its speed in the low corona, where the Lorentz force is stronger, was certainly lower than that in our simulation. As noted above, the simulation does not reproduce the slow rise of the CME for the 18 minutes from the start of the SXR flare to the start of the HXR flare. It might create an additional deflection of about 5--10$^\circ$ in the low corona, assuming that the actual deflection rate is more or less consistent with what we find in our simulation. Another reason why our simulation does not reproduce the observed deflection is also due the initiation mechanism. As the force-free field of the flux rope is superimposed onto the coronal magnetic field, it results in a magnetic configuration which is not force-free. Therefore, there are additional Lorentz forces on all directions, which partially compensate the main eastward Lorentz force discussed before. In the real Sun, it is likely that the filament was in a force-free state before the eruption and it would only be subject to the main eastward Lorentz force, which would yield a larger deflection. Deflection of CMEs from the same active region during the following Carrington rotation has also been reported \citep[]{Wang:2006}.

\section{DISCUSSIONS AND CONCLUSION}\label{sec:conclusion}

We have investigated, using a numerical simulation, the first of two eruptions on 2005 August 22 from anemone active region 10798. We have used a recently developed component of the SWMF to include realistic thermodynamics into the representation of the low corona (LC). Using the LC component has enabled us to produce EUV images of the solar corona which we can compare to EIT observations. In addition, previous works have shown that including thermodynamics is important to reproduce the plasma and magnetic properties of the entire corona. We have investigated the pre-event topology of the active region and found that it indeed developed into an anemone active region with a single null point separating the active region flux system to unipolar positive magnetic field. We have found that some of these unipolar magnetic field lines are open while others are part of the streamer belt connected to a region near disk center. Pre-event 171~\AA~and 195~\AA~images of the corona are in good agreement with simulated ones, validating our steady-state model. In agreement with observations of a H$\alpha$ filament, which was well observed prior to the eruption, we have superposed a highly twisted flux rope onto the steady-state solar corona at the same location and with the same orientation and chirality (right-handed and southward). Due to force imbalance, the flux rope immediately erupts and we have focused our study on its interaction with the adjacent magnetic flux systems and in particular with the open flux from the surrounding coronal holes. 

The anemone nature of the source region has two major consequences on the evolution of the CME: an eastward deflection due to the Lorentz force and a reconnection of the negative footpoint of the flux rope, which results in the appearance of a long-duration dimming region and a mix of open and closed field lines in the CME. Since the flux rope was southward and right-handed, there was a strong southward axial current. The Lorentz force generated by this current and the unipolar positive magnetic field in the vicinity of the active region is east-directed and, in our simulation, it deflects the CME by about 10--15$^\circ$ as the CME propagates to 8~R$_\odot$. We have not address the reason why CMEs from anemone active regions tend to be fast CMEs \citep[]{Liu:2007}. It should be however noted that, in the simulation, the CME encounters closed field lines in the low corona, so the magnetic topology cannot be fully responsible for the fast speed. It is therefore likely that the fast speed of CMEs from anemone active regions is at least partially due to the fact the CME propagates into a low density, fast solar wind, where the hydrodynamical drag is low \citep[]{Cargill:2004}.

The negative footpoint of the flux rope quickly reconnects with the positive field part of the streamer belt. It leads to a CME with one footpoint (the positive one) that is line-tied, and the second one which connects to a region of the Sun about 40$^\circ$ away from the active region. A dimming region was found to develop in the region of the second footpoint, both in simulated EUV images and in EIT images and to persist at the Sun for as much as a day after the eruption. Additionally, some of these closed field lines reconnect with open field from the equatorial coronal hole in a type of interchange reconnection previously discussed in \citet{Baker:2009}. It yields a mix of open and closed field lines in the erupting ejecta. While we have conducted our simulation only to 10~R$_\odot$, this type of mixed nature of the erupting field is expected from observations of periods of unidirectional strahl inside periods of bi-directional electrons as a magnetic cloud passes over a spacecraft \citep[]{Gosling:1995, Shodhan:2000}. To the best of our knowledge, this is the first time such a mixture is obtained from a numerical simulation of a real CME event. The fact that we obtain interchange reconnection in our simulation is not surprising, since it has been noted before that anemone active regions are the most suitable environment for interchange reconnection \citep[]{Baker:2009}.

We have also found that the CME does not rotate as it propagates in the corona (up to a distance of 10~R$_\odot$). This result is found although there is large-scale reconnection, the CME axis is initially not aligned with the HCS and the flux rope deforms into a skewed tube at a distance of less than 1.7~R$_\odot$. It shows that the rotation of CME axis is not a universal process and that some CMEs maintain their orientation in spite of large-scale reconnection. This result is a pendant to previous numerical studies \citep[]{Lynch:2009, Cohen:2010, Shiota:2010} which found that rotation happens when a CME is initially misaligned with the HCS, or when it is skewed. The deflection of the CME, the persistence of closed field lines into the inner heliosphere and the lack of rotation are also consistent with the observation of a short magnetic cloud at 1~AU about 2 days after the first eruption. \citet{Asai:2009} associated this magnetic cloud with the first of the two eruptions of 2005 August 22. 

While our simulation captures some of the most important observational features of the CME, there are some discrepancies. In our simulation, the CME does appear to propagate along the same latitude as that of its source region (S11) while LASCO images show a CME propagating close to the ecliptic. One likely explanation is that only the northern part of the filament erupted \citep[as reported in][]{Asai:2009}. Our method to obtain an eruption is too simple to study partial filament eruption or to reproduce the exact dynamics and kinematics of the eruption at distances less than 3~R$_\odot$. Another difference between the simulation and the observations is the amount of deflection. In our simulation, the CME deflects by about 10$^\circ$ eastward whereas a total deflection of about 40--50$^\circ$ is expected from observations. A smaller deflection might be present in our simulation compared to what happened on the real Sun because of different kinematics and force balance in the low corona. However, we believe that a significant portion of the deflection cannot be accounted by magnetic forces, which are dominant in the low corona. This finding implies that a significant part of the deflection is due to hydrodynamical effect. For this CME from an anemone active region, followed by a twice faster ejection 18 hours later, these effects may take two forms: deflection by the fast wind from the coronal hole as discussed, for example, in \citet{Wang:2004} and \citet{Gopalswamy:2009} and also deflection by the second, faster CME pushing the eruption from behind, as discussed in \citet{Xiong:2009}. In this case, the additional deflection can be expected to happen more gradually in the heliosphere. Future simulations will investigate the heliospheric evolution of CMEs from anemone active regions and overtaken by faster eruption.

\begin{acknowledgments}
The research for this manuscript was performed during and supported by a JSPS short-term post-doctoral fellowship at Kwasan Observatory in Kyoto University. We would like to thank the reviewer and Terry Forbes for their helpful comments, which helped us clarify the discussion of the reconnection process. N.~L. would like to thank everyone from the Kwasan Observatory for their welcome during his 7-month stay there. N.~L. and I.~R. acknowledge additional financial support from NSF grants ATM0639335 and ATM0819653 and NASA grant NNX08AQ16G. C.~D. was supported by a NASA NESSF08-Helio08F-0007.
\end{acknowledgments}

\bibliographystyle{apj}


\begin{thebibliography}{66}
\expandafter\ifx\csname natexlab\endcsname\relax\def\natexlab#1{#1}\fi

\bibitem[{{Altschuler} {et~al.}(1977){Altschuler}, {Levine}, {Stix}, \&
  {Harvey}}]{Altschuler:1977}
{Altschuler}, M.~D., {Levine}, R.~H., {Stix}, M., \& {Harvey}, J. 1977, Solar
  Phys., 51, 345

\bibitem[{{Amari} {et~al.}(2010){Amari}, {Aly}, {Mikic}, \&
  {Linker}}]{Amari:2010}
{Amari}, T., {Aly}, J., {Mikic}, Z., \& {Linker}, J. 2010, Astrophys. Journ.
  Lett., 717, L26

\bibitem[{{Asai} {et~al.}(2008){Asai}, {Shibata}, {Hara}, \&
  {Nitta}}]{Asai:2008}
{Asai}, A., {Shibata}, K., {Hara}, H., \& {Nitta}, N.~V. 2008, Astrophys. J.,
  673, 1188

\bibitem[{{Asai} {et~al.}(2009){Asai}, {Shibata}, {Ishii}, {Oka}, {Kataoka},
  {Fujiki}, \& {Gopalswamy}}]{Asai:2009}
{Asai}, A., {Shibata}, K., {Ishii}, T.~T., {Oka}, M., {Kataoka}, R., {Fujiki},
  K., \& {Gopalswamy}, N. 2009, J. Geophys. Res., 114, {A00A21}

\bibitem[Attrill et al.(2006)]{Attrill:2006} Attrill, G., Nakwacki, 
M.~S., Harra, L.~K., van Driel-Gesztelyi, L., Mandrini, C.~H., Dasso, S., 
\& Wang, J.\ 2006,  Solar Phys., 238, 117 

\bibitem[Attrill et al.(2007)]{Attrill:2007} Attrill, G.~D.~R., 
Harra, L.~K., van Driel-Gesztelyi, L., 
\& D{\'e}moulin, P.\ 2007, Astrophys. Journ.
  Lett., 656, L101 

\bibitem[{{Aulanier} {et~al.}(2010){Aulanier}, {T{\"o}r{\"o}k}, {D{\'e}moulin},
  \& {DeLuca}}]{Aulanier:2010}
{Aulanier}, G., {T{\"o}r{\"o}k}, T., {D{\'e}moulin}, P., \& {DeLuca}, E.~E.
  2010, Astrophys. J., 708, 314

\bibitem[Baker et al.(2009)]{Baker:2009} Baker, D., et al.\ 2009, 
Annales Geophysicae, 27, 3883 

\bibitem[{{Byrne} {et~al.}(2010){Byrne}, {Maloney}, {McAteer}, {Refojo}, \&
  {Gallagher}}]{Byrne:2010}
{Byrne}, J.~P., {Maloney}, S.~A., {McAteer}, R.~T.~J., {Refojo}, J.~M., \&
  {Gallagher}, P.~T. 2010, Nature Communications, 1

\bibitem[Cargill(2004)]{Cargill:2004} Cargill, P.~J.\ 2004, 
\solphys, 221, 135 

\bibitem[{{Chen}(1996)}]{Chen:1996}
{Chen}, J. 1996, J. Geophys. Res., 101, 27499

\bibitem[Chen 
\& Shibata(2000)]{Chen:2000} Chen, P.~F., \& Shibata, K.\ 2000, \apj, 545, 52

\bibitem[{{Chertok} {et~al.}(2002){Chertok}, {Obridko}, {Mogilevsky},
  {Shilova}, \& {Hudson}}]{Chertok:2002}
{Chertok}, I.~M., {Obridko}, E.~I., {Mogilevsky}, V.~N., {Shilova}, N.~S., \&
  {Hudson}, H.~S. 2002, Astrophys. J., 567, 1225

\bibitem[{{Cohen} {et~al.}(2010){Cohen}, {Attrill}, {Schwadron}, {Crooker},
  {Owens}, {Downs}, \& {Gombosi}}]{Cohen:2010}
{Cohen}, O., {Attrill}, G.~D.~R., {Schwadron}, N.~A., {Crooker}, N.~U.,
  {Owens}, M.~J., {Downs}, C., \& {Gombosi}, T.~I. 2010, J. Geophys. Res., 115

\bibitem[{{Cohen} {et~al.}(2007){Cohen}, {Sokolov}, {Roussev}, {Arge},
  {Manchester}, {Gombosi}, {Frazin}, {Park}, {Butala}, {Kamalabadi}, \&
  {Velli}}]{Cohen:2007}
{Cohen}, O., {et~al.} 2007, Astrophys. Journ. Lett., 654, L163

\bibitem[{{Cremades} {et~al.}(2006){Cremades}, {Bothmer}, \&
  {Tripathi}}]{Cremades:2006}
{Cremades}, H., {Bothmer}, V., \& {Tripathi}, D. 2006, Adv. Space Res., 38, 461

\bibitem[Crooker et al.(2002)]{Crooker:2002} Crooker, N.~U., 
Gosling, J.~T., 
\& Kahler, S.~W.\ 2002, J. Geophys. Res., 107, 1028 

\bibitem[{{Delaboudini{\`e}re} {et~al.}(1995){Delaboudini{\`e}re}, {Artzner},
  {Brunaud}, {Gabriel}, {Hochedez}, {Millier}, {Song}, {Au}, {Dere}, {Howard},
  {Kreplin}, {Michels}, {Moses}, {Defise}, {Jamar}, {Rochus}, {Chauvineau},
  {Marioge}, {Catura}, {Lemen}, {Shing}, {Stern}, {Gurman}, {Neupert},
  {Maucherat}, {Clette}, {Cugnon}, \& {van Dessel}}]{Delaboudiniere:1995}
{Delaboudini{\`e}re}, J., {et~al.} 1995, Solar Phys., 162, 291

\bibitem[{{Downs} {et~al.}(2010){Downs}, {Roussev}, {van der Holst}, {Lugaz},
  {Sokolov}, \& {Gombosi}}]{Downs:2010}
{Downs}, C., {Roussev}, I.~I., {van der Holst}, B., {Lugaz}, N., {Sokolov},
  I.~V., \& {Gombosi}, T.~I. 2010, Astrophys. J., 712, 1219

\bibitem[{{Downs} {et~al.}(2011){Downs}, {Roussev}, {van der Holst}, {Lugaz},
  {Sokolov}, \& {Gombosi}}]{Downs:2011}
---. 2011, Astrophys. J., 728, 2

\bibitem[Edmondson et al.(2010)]{Edmondson:2010} Edmondson, J.~K., 
Antiochos, S.~K., DeVore, C.~R., Lynch, B.~J., 
\& Zurbuchen, T.~H.\ 2010, \apj, 714, 517 

\bibitem[{{Evans} {et~al.}(2011){Evans}, {Opher}, \& {Gombosi}}]{Evans:2011}
{Evans}, R.~M., {Opher}, M., \& {Gombosi}, T.~I. 2011, Astrophys. J., 728, 41

\bibitem[{{Fan}(2005)}]{Fan:2005}
{Fan}, Y. 2005, Astrophys. J., 630, 543

\bibitem[Filippov et al.(2001)]{Filippov:2001} Filippov, B.~P., 
Gopalswamy, N., \& Lozhechkin, A.~V.\ 2001, \solphys, 203, 119 

\bibitem[{{Forbes}(2000)}]{Forbes:2000}
{Forbes}, T.~G. 2000, J. Geophys. Res., 105, 23153

\bibitem[{{Gopalswamy} {et~al.}(2009){Gopalswamy}, {M{\"a}kel{\"a}}, {Xie},
  {Akiyama}, \& {Yashiro}}]{Gopalswamy:2009}
{Gopalswamy}, N., {M{\"a}kel{\"a}}, P., {Xie}, H., {Akiyama}, S., \& {Yashiro},
  S. 2009, J. Geophys. Res., 114, {A00A22}

\bibitem[Gosling et al.(1995)]{Gosling:1995} Gosling, J.~T., Birn, 
J., \& Hesse, M.\ 1995, \grl, 22, 869 

\bibitem[{{Hale} {et~al.}(1919){Hale}, {Ellerman}, {Nicholson}, \&
  {Joy}}]{Hale:1919}
{Hale}, G.~E., {Ellerman}, F., {Nicholson}, S.~B., \& {Joy}, A.~H. 1919,
  Astrophys. J., 49, 153

\bibitem[{{Hill} {et~al.}(2005){Hill}, {Pizzo}, {Balch}, {Biesecker},
  {Bornmann}, {Hildner}, {Lewis}, {Grubb}, {Husler}, {Prendergast}, {Vickroy},
  {Greer}, {Defoor}, {Wilkinson}, {Hooker}, {Mulligan}, {Chipman}, {Bysal},
  {Douglas}, {Reynolds}, {Davis}, {Wallace}, {Russell}, {Freestone},
  {Bagdigian}, {Page}, {Kerns}, {Hoffman}, {Cauffman}, {Davis}, {Studer},
  {Berthiaume}, {Saha}, {Berthiume}, {Farthing}, \& {Zimmermann}}]{Hill:2005}
{Hill}, S.~M., {et~al.} 2005, Solar Phys., 226, 255

\bibitem[{{Isenberg} \& {Forbes}(2007)}]{Isenberg:2007}
{Isenberg}, P.~A., \& {Forbes}, T.~G. 2007, Astrophys. J., 670, 1453

\bibitem[{{Jacobs} {et~al.}(2009){Jacobs}, {Roussev}, {Lugaz}, \&
  {Poedts}}]{Jacobs:2009}
{Jacobs}, C., {Roussev}, I.~I., {Lugaz}, N., \& {Poedts}, S. 2009, Astrophys.
  Journ. Lett., 695, L171

\bibitem[{{Kilpua} {et~al.}(2008){Kilpua}, {Liewer}, {Farrugia}, {Luhmann},
  {M{\"o}stl}, {Li}, {Liu}, {Lynch}, {Russell}, {Vourlidas}, {Acuna}, {Galvin},
  {Larson}, \& {Sauvaud}}]{Kilpua:2009a}
{Kilpua}, E.~K.~J., {et~al.} 2008, Solar Phys., 254, 325

\bibitem[{{Lin} {et~al.}(2002){Lin}, {Dennis}, {Hurford}, {Smith}, {Zehnder},
  {Harvey}, {Curtis}, {Pankow}, {Turin}, {Bester}, {Csillaghy}, {Lewis},
  {Madden}, {van Beek}, {Appleby}, {Raudorf}, {McTiernan}, {Ramaty}, {Schmahl},
  {Schwartz}, {Krucker}, {Abiad}, {Quinn}, {Berg}, {Hashii}, {Sterling},
  {Jackson}, {Pratt}, {Campbell}, {Malone}, {Landis}, {Barrington-Leigh},
  {Slassi-Sennou}, {Cork}, {Clark}, {Amato}, {Orwig}, {Boyle}, {Banks},
  {Shirey}, {Tolbert}, {Zarro}, {Snow}, {Thomsen}, {Henneck}, {McHedlishvili},
  {Ming}, {Fivian}, {Jordan}, {Wanner}, {Crubb}, {Preble}, {Matranga}, {Benz},
  {Hudson}, {Canfield}, {Holman}, {Crannell}, {Kosugi}, {Emslie}, {Vilmer},
  {Brown}, {Johns-Krull}, {Aschwanden}, {Metcalf}, \& {Conway}}]{Lin:2002}
{Lin}, R.~P., {et~al.} 2002, Solar Phys., 210, 3

\bibitem[{{Lionello} {et~al.}(2001){Lionello}, {Linker}, \&
  {Miki{\'c}}}]{Lionello:2001}
{Lionello}, R., {Linker}, J.~A., \& {Miki{\'c}}, Z. 2001, Astrophys. J., 546,
  542

\bibitem[{{Lionello} {et~al.}(2009){Lionello}, {Linker}, \&
  {Miki{\'c}}}]{Lionello:2009}
---. 2009, Astrophys. J., 690, 902

\bibitem[{{Liu}(2007)}]{Liu:2007}
{Liu}, Y. 2007, Astrophys. Journ. Lett., 654, L171

\bibitem[{{Liu} {et~al.}(2010){Liu}, {Thernisien}, {Luhmann}, {Vourlidas},
  {Davies}, {Lin}, \& {Bale}}]{Liu:2010b}
{Liu}, Y., {Thernisien}, A., {Luhmann}, J.~G., {Vourlidas}, A., {Davies},
  J.~A., {Lin}, R.~P., \& {Bale}, S.~D. 2010, Astrophys. J., 722, 1762

\bibitem[{{Lugaz} {et~al.}(2005){Lugaz}, {Manchester}, \&
  {Gombosi}}]{Lugaz:2005b}
{Lugaz}, N., {Manchester}, W.~B., \& {Gombosi}, T.~I. 2005, Astrophys. J., 634,
  651

\bibitem[{{Lugaz} {et~al.}(2007){Lugaz}, {Manchester}, {Roussev}, {T{\'o}th},
  \& {Gombosi}}]{Lugaz:2007}
{Lugaz}, N., {Manchester}, W.~B., {Roussev}, I.~I., {T{\'o}th}, G., \&
  {Gombosi}, T.~I. 2007, Astrophys. J., 659, 788

\bibitem[{{Lugaz} {et~al.}(2010){Lugaz}, {Roussev}, {Sokolov}, \&
  {Jacobs}}]{Lugaz:2010a}
{Lugaz}, N., {Roussev}, I.~I., {Sokolov}, I.~V., \& {Jacobs}, C. 2010, Twelfth
  International Solar Wind Conference, 1216, 440

\bibitem[{{Lugaz} {et~al.}(2009){Lugaz}, {Vourlidas}, {Roussev}, \&
  {Morgan}}]{Lugaz:2009b}
{Lugaz}, N., {Vourlidas}, A., {Roussev}, I.~I., \& {Morgan}, H. 2009, Solar
  Phys., 256, 269

\bibitem[{{Lynch} {et~al.}(2009){Lynch}, {Antiochos}, {Li}, {Luhmann}, \&
  {DeVore}}]{Lynch:2009}
{Lynch}, B.~J., {Antiochos}, S.~K., {Li}, Y., {Luhmann}, J.~G., \& {DeVore},
  C.~R. 2009, Astrophys. J., 697, 1918

\bibitem[Mackay 
\& van Ballegooijen(2006)]{MacKay:2006} Mackay, D.~H., \& van Ballegooijen, A.~A.\ 2006, \apj, 642, 1193 

\bibitem[Mackay et al.(2010)]{MacKay:2010} Mackay, D.~H., Karpen, 
J.~T., Ballester, J.~L., Schmieder, B., \& Aulanier, G.\ 2010, Space Sci. Rev., 151, 333 

\bibitem[{{MacQueen} {et~al.}(1986){MacQueen}, {Hundhausen}, \&
  {Conover}}]{MacQueen:1986}
{MacQueen}, R.~M., {Hundhausen}, A.~J., \& {Conover}, C.~W. 1986, J. Geophys.
  Res., 91, 31

\bibitem[{{Manchester}(2003)}]{Manchester:2003}
{Manchester}, W.~B. 2003, J. Geophys. Res., 108, 10

\bibitem[{{Manchester} {et~al.}(2004){Manchester}, {Gombosi}, {DeZeeuw}, \&
  {Fan}}]{Manchester:2004c}
{Manchester}, W.~B., {Gombosi}, T., {DeZeeuw}, D., \& {Fan}, Y. 2004,
  Astrophys. J., 610, 588

\bibitem[Mandrini et al.(2007)]{Mandrini:2007} Mandrini, C.~H., 
Nakwacki, M.~S., Attrill, G., van Driel-Gesztelyi, L., D{\'e}moulin, P., 
Dasso, S., \& Elliott, H.\ 2007, Solar Phys., 244, 25 

\bibitem[{{Morgan} {et~al.}(2006){Morgan}, {Habbal}, \& {Woo}}]{Morgan:2006}
{Morgan}, H., {Habbal}, S.~R., \& {Woo}, R. 2006, Solar Phys., 236, 263

\bibitem[{{Ogawara} {et~al.}(1991){Ogawara}, {Takano}, {Kato}, {Kosugi},
  {Tsuneta}, {Watanabe}, {Kondo}, \& {Uchida}}]{Ogawara:1991}
{Ogawara}, Y., {Takano}, T., {Kato}, T., {Kosugi}, T., {Tsuneta}, S.,
  {Watanabe}, T., {Kondo}, I., \& {Uchida}, Y. 1991, Solar Phys., 136, 1

\bibitem[{{Ohyama} \& {Shibata}(1998)}]{Ohyama:1998}
{Ohyama}, M., \& {Shibata}, K. 1998, Astrophys. J., 499, 934

\bibitem[Pariat et al.(2009)]{Pariat:2009} Pariat, E., Antiochos, 
S.~K., \& DeVore, C.~R.\ 2009, Astrophys. J., 691, 61 

\bibitem[Parnell 
\& Galsgaard(2004)]{Parnell:2004} Parnell, C.~E., \& Galsgaard, K.\ 2004, \aap, 428, 595 

\bibitem[{{Plunkett} {et~al.}(2001){Plunkett}, {Thompson}, {St.~Cyr}, \&
  {Howard}}]{Plunkett:2001}
{Plunkett}, S.~P., {Thompson}, B.~J., {St.~Cyr}, O.~C., \& {Howard}, R.~A.
  2001, J. Atmos. Solar-Terr. Phys., 63, 389

\bibitem[{{Roussev} {et~al.}(2003){Roussev}, {Forbes}, {Gombosi}, {Sokolov},
  {DeZeeuw}, \& {Birn}}]{Roussev:2003a}
{Roussev}, I.~I., {Forbes}, T.~G., {Gombosi}, T.~I., {Sokolov}, I.~V.,
  {DeZeeuw}, D.~L., \& {Birn}, J. 2003, Astrophys. Journ. Lett., 588, L45

\bibitem[{{Roussev} {et~al.}(2007){Roussev}, {Lugaz}, \&
  {Sokolov}}]{Roussev:2007}
{Roussev}, I.~I., {Lugaz}, N., \& {Sokolov}, I.~V. 2007, Astrophys. Journ.
  Lett., 668, L87

\bibitem[{{Sheeley} {et~al.}(1975){Sheeley}, {Bohlin}, {Brueckner}, {Purcell},
  {Scherrer}, \& {Tousey}}]{Sheeley:1975}
{Sheeley}, Jr., N.~R., {Bohlin}, J.~D., {Brueckner}, G.~E., {Purcell}, J.~D.,
  {Scherrer}, V., \& {Tousey}, R. 1975, Solar Phys., 40, 103

\bibitem[{{Shen} {et~al.}(2011){Shen}, {Wang}, {Gui}, {Ye}, \&
  {Wang}}]{Shen:2011}
{Shen}, C., {Wang}, Y., {Gui}, B., {Ye}, P., \& {Wang}, S. 2011, Solar Phys.,
269, 389 

\bibitem[{{Shibata} {et~al.}(1994){Shibata}, {Nitta}, {Strong}, {Matsumoto},
  {Yokoyama}, {Hirayama}, {Hudson}, \& {Ogawara}}]{Shibata:1994}
{Shibata}, K., {Nitta}, N., {Strong}, K.~T., {Matsumoto}, R., {Yokoyama}, T.,
  {Hirayama}, T., {Hudson}, H., \& {Ogawara}, Y. 1994, Astrophys. Journ. Lett.,
  431, L51

\bibitem[{{Shibata} {et~al.}(2007){Shibata}, {Nakamura}, {Matsumoto}, {Otsuji},
  {Okamoto}, {Nishizuka}, {Kawate}, {Watanabe}, {Nagata}, {UeNo}, {Kitai},
  {Nozawa}, {Tsuneta}, {Suematsu}, {Ichimoto}, {Shimizu}, {Katsukawa},
  {Tarbell}, {Berger}, {Lites}, {Shine}, \& {Title}}]{Shibata:2007}
{Shibata}, K., {et~al.} 2007, Science, 318, 1591

\bibitem[{{Shiota} {et~al.}(2010){Shiota}, {Kusano}, {Miyoshi}, \&
  {Shibata}}]{Shiota:2010}
{Shiota}, D., {Kusano}, K., {Miyoshi}, T., \& {Shibata}, K. 2010, Astrophys.
  J., 718, 1305

\bibitem[Shodhan et al.(2000)]{Shodhan:2000} Shodhan, S., Crooker, 
N.~U., Kahler, S.~W., Fitzenreiter, R.~J., Larson, D.~E., Lepping, R.~P., 
Siscoe, G.~L., \& Gosling, J.~T.\ 2000, J. Geophys. Res., 105, 27261 

\bibitem[{{Temmer} {et~al.}(2008){Temmer}, {Veronig}, {Vr{\v s}nak},
  {Ryb{\'a}k}, {G{\"o}m{\"o}ry}, {Stoiser}, \& {Mari{\v
  c}i{\'c}}}]{Temmer:2008}
{Temmer}, M., {Veronig}, A.~M., {Vr{\v s}nak}, B., {Ryb{\'a}k}, J.,
  {G{\"o}m{\"o}ry}, P., {Stoiser}, S., \& {Mari{\v c}i{\'c}}, D. 2008,
  Astrophys. Journ. Lett., 673, L95

\bibitem[{{Titov} \& {D{\'e}moulin}(1999)}]{Titov:1999}
{Titov}, V.~S., \& {D{\'e}moulin}, P. 1999, Astron. Astrophys., 351, 707

\bibitem[{{Titov} {et~al.}(2008){Titov}, {Mikic}, {Linker}, \&
  {Lionello}}]{Titov:2008}
{Titov}, V.~S., {Mikic}, Z., {Linker}, J.~A., \& {Lionello}, R. 2008,
  Astrophys. J., 675, 1614

\bibitem[{{T{\"o}r{\"o}k} \& {Kliem}(2003)}]{Torok:2003}
{T{\"o}r{\"o}k}, T., \& {Kliem}, B. 2003, Astron. Astrophys., 406, 1043

\bibitem[{{T{\'o}th} {et~al.}(2005){T{\'o}th}, {Sokolov}, {Gombosi}, {Chesney},
  {Clauer}, {De Zeeuw}, {Hansen}, {Kane}, {Manchester}, {Oehmke}, {Powell},
  {Ridley}, {Roussev}, {Stout}, {Volberg}, {Wolf}, {Sazykin}, {Chan}, {Yu}, \&
  {K{\'o}ta}}]{Toth:2005}
{T{\'o}th}, G., {et~al.} 2005, J. Geophys. Res., 110, {A12226}

\bibitem[T{\'o}th et al.(2011)]{Toth:2011} T{\'o}th, G., van der 
Holst, B., \& Huang, Z.\ 2011, Astrophys. J., 732, 102 

\bibitem[{{Tsuneta} {et~al.}(1991){Tsuneta}, {Acton}, {Bruner}, {Lemen},
  {Brown}, {Caravalho}, {Catura}, {Freeland}, {Jurcevich}, \&
  {Owens}}]{Tsuneta:1991}
{Tsuneta}, S., {et~al.} 1991, Solar Phys., 136, 37

\bibitem[{{van der Holst} {et~al.}(2010){van der Holst}, {Manchester},
  {Frazin}, {V{\'a}squez}, {T{\'o}th}, \& {Gombosi}}]{Holst:2010}
{van der Holst}, B., {Manchester}, W.~B., {Frazin}, R.~A., {V{\'a}squez},
  A.~M., {T{\'o}th}, G., \& {Gombosi}, T.~I. 2010, Astrophys. J., 725, 1373

\bibitem[{{Vourlidas} {et~al.}(1996){Vourlidas}, {Bastian}, {Nitta}, \&
  {Aschwanden}}]{Vourlidas:1996}
{Vourlidas}, A., {Bastian}, T.~S., {Nitta}, N., \& {Aschwanden}, M.~J. 1996,
  Solar Phys., 163, 99

\bibitem[{{Wang} {et~al.}(2004){Wang}, {Shen}, {Wang}, \& {Ye}}]{Wang:2004}
{Wang}, Y., {Shen}, C., {Wang}, S., \& {Ye}, P. 2004, Solar Phys., 222, 329

\bibitem[Wang et al.(2006)]{Wang:2006} Wang, Y., Xue, X., Shen, 
C., Ye, P., Wang, S., \& Zhang, J.\ 2006, Astrophys. J., 646, 625 

\bibitem[Webb et al.(2000)]{Webb:2000} Webb, D.~F., Lepping, 
R.~P., Burlaga, L.~F., DeForest, C.~E., Larson, D.~E., Martin, S.~F., 
Plunkett, S.~P., \& Rust, D.~M.\ 2000,  J. Geophys. Res., 105, 27251 

\bibitem[{{Xiong} {et~al.}(2009){Xiong}, {Zheng}, \& {Wang}}]{Xiong:2009}
{Xiong}, M., {Zheng}, H., \& {Wang}, S. 2009, J. Geophys. Res., 114, A11101

\bibitem[{{Yurchyshyn}(2008)}]{Yurchyshyn:2008}
{Yurchyshyn}, V. 2008, Astrophys. Journ. Lett., 675, L49

\bibitem[{{Yurchyshyn} {et~al.}(2009){Yurchyshyn}, {Abramenko}, \&
  {Tripathi}}]{Yurchyshyn:2009}
{Yurchyshyn}, V., {Abramenko}, V., \& {Tripathi}, D. 2009, Astrophys. J., 705,
  426

\end{thebibliography}


\end{document}